\documentclass[prd,%
preprint,%
tightenlines,showpacs,%
nofootinbib,%
superscriptaddress,%
amsfonts,amsmath,%
floatfix]{revtex4}

\usepackage{hyperref}
\usepackage{graphicx}

\begin{document}

\title{Hamiltonian unboundedness vs stability\\
with an application to Horndeski theory}

\author{E.~Babichev}
\email{eugeny.babichev@th.u-psud.fr}
\affiliation{Laboratoire de Physique Th\'eorique,
CNRS, Univ. Paris-Sud,\\
Universit\'e Paris-Saclay, F-91405 Orsay, France}
\affiliation{Sorbonne Universit\'e, CNRS, UMR7095,
Institut d'Astrophysique de Paris,
${\mathcal{G}}{\mathbb{R}}\varepsilon{\mathbb{C}}{\mathcal{O}}$,\\
98bis boulevard Arago, F-75014 Paris, France}

\author{C.~Charmousis}
\email{christos.charmousis@th.u-psud.fr}
\affiliation{Laboratoire de Physique Th\'eorique,
CNRS, Univ. Paris-Sud,\\
Universit\'e Paris-Saclay, F-91405 Orsay, France}

\author{G.~Esposito-Far\`ese}
\email{gef@iap.fr}
\affiliation{Sorbonne Universit\'e, CNRS, UMR7095,
Institut d'Astrophysique de Paris,
${\mathcal{G}}{\mathbb{R}}\varepsilon{\mathbb{C}}{\mathcal{O}}$,\\
98bis boulevard Arago, F-75014 Paris, France}

\author{A.~Leh\'ebel}
\email{antoine.lehebel@th.u-psud.fr}
\affiliation{Laboratoire de Physique Th\'eorique,
CNRS, Univ. Paris-Sud,\\
Universit\'e Paris-Saclay, F-91405 Orsay, France}

\begin{abstract}A Hamiltonian density bounded from below implies
that the lowest-energy state is stable. We point out, contrary to
common lore, that an unbounded Hamiltonian density does not
necessarily imply an instability: Stability is indeed a
coordinate-independent property, whereas the Hamiltonian density
does depend on the choice of coordinates. We discuss in detail
the relation between the two, starting from k-essence and
extending our discussion to general field theories. We give the
correct stability criterion, using the relative orientation of
the causal cones for all propagating degrees of freedom. We then
apply this criterion to an exact Schwarzschild-de Sitter solution
of a beyond-Horndeski theory, while taking into account the
recent experimental constraint regarding the speed of
gravitational waves. We extract the spin-2 and spin-0 causal
cones by analyzing respectively all the odd-parity and the
$\ell=0$ even-parity modes. Contrary to a claim in the
literature, we prove that this solution does not exhibit any
kinetic instability for a given range of parameters defining the
theory.
\end{abstract}

\pacs{04.50.Kd, 98.80.-k, 04.70.Bw}
\date{September 10, 2018}

\maketitle

\section{Introduction}

General relativity (GR) is an effective classical theory of
gravity which is experimentally verified for a wide range of
physical distance and gravitational strength scales. The former
range up to 30 astronomical units or so, while the latter are
performed from weak gravity tabletop experiments to very strong
gravity environments of coalescing black hole binaries, involving
recent gravitational wave detections (see \cite{Will:2014kxa} for
a review, and for the latter
\cite{TheLIGOScientific:2017qsa,Abbott:2016blz}). GR is
furthermore a theoretically robust theory, as it is unique when
one imposes standard mathematical and physical assumptions:
essentially the presence of a Levi-Civita connection for a
sufficiently regular four-dimensional spacetime manifold equipped
with a metric mediating gravitational interactions. These dictate
that GR with a cosmological constant is the unique covariant
theory with second-order field equations \cite{Lovelock:1971yv}.
Since no mass term is associated with the rank-2 tensor mediating
gravity, GR has two massless spin-2 degrees of freedom. A second
and independent assumption of GR is that all matter fields
universally couple to this metric, in order to satisfy the weak
equivalence principle. These postulates imply that photons and
gravitons propagate on the same causal cones, i.e., with the same
speed.

The presence of dark energy at cosmological distance scales (but
also the yet elusive dark matter) has opened up in recent years
the possibility that GR may be an effective theory not only at UV
scales but also at the deep IR: cosmological ---~but also lack of
astrophysical~--- observations have raised questions concerning
the viability of GR at large distances (see for example
\cite{Clifton:2011jh}). Given the aforementioned uniqueness, if
we want to go beyond GR, we have to introduce novel degrees of
freedom. This is true if we remain within the realm of Riemannian
geometry, keeping the postulated geometrical structure of
spacetime. The simplest of these degrees of freedom is a scalar
field, giving scalar-tensor theories of gravity (see
\cite{Will:1993ns,Damour:1992we,Horndeski:1974wa,Fairlie,Nicolis:%
2008in,Deffayet:2009wt,Deffayet:2009mn,deRham:2010eu,Deffayet:%
2011gz,Deffayet:2013lga}), but one can also consider vector(s)
(see for example
\cite{Horndeski:1976gi,Deffayet:2010zh,Heisenberg:2014rta,%
Tasinato:2014eka,Kase:2018owh}), an additional metric field (see
\cite{deRham:2014zqa} for a review), etc. One should point out
that even more general considerations, such as the presence of
non-trivial torsion (e.g. \cite{Zanelli:2005sa}) or spacetime
with extra dimensions (see \cite{Charmousis:2017rof} and
references therein), also lead effectively to the addition of
extra degrees of freedom. The prototype example is that of the
Dvali-Gabadadze-Porrati (DGP) five-dimensional\footnote{See also
\cite{Charmousis:2014mia} for relations between higher
dimensional and 4-dimensional scalar-tensor gravity theories.}
braneworld model \cite{Dvali:2000hr}, which upon going to the
decoupling limit \cite{Nicolis:2004qq} reduces to a particular
Horndeski scalar-tensor theory (of the type studied in
\cite{Babichev:2012re}).

Modified gravity degrees of freedom ---~including gravitons~---
propagate in an effective metric which can be different from that
of GR. The notion of causal cones and effective metric will be
essential to our stability arguments, so let us dwell on this
point before entering details in the body of the paper. Consider
some background solution of a modified gravity theory. Linear
perturbations of the background can be found by expanding the
action up to the second order.
According to their effective action, modes obey some second-order
differential equation\footnote{We do not consider here
Lorentz-breaking theories \cite{Horava:2009uw,Jacobson:2008aj}
where equations of motion can be of higher order. Also, in
extensions of Horndeski theory
\cite{Zumalacarregui:2013pma,Gleyzes:2014dya,%
Lin:2014jga,Gleyzes:2014qga,Deffayet:2015qwa,Langlois:2015cwa,%
Langlois:2015skt,Crisostomi:2016tcp,Crisostomi:2016czh,%
Achour:2016rkg,deRham:2016wji,BenAchour:2016fzp,%
Motohashi:2014opa,Motohashi:2016ftl},
one may get Euler-Lagrange equations of third order, but by
manipulating these equations their order is reduced.}.
We will concentrate on the kinetic part of the equations of
motion (i.e., the principal part of the differential equation),
since this part defines whether the most dangerous pathologies of
the theory are absent. The kinetic operator is encoded in an
effective metric, and in general it depends on the background
solution.
Provided that the equation of motion is hyperbolic, the kinetic
operator (i.e., the effective metric) defines then a causal cone
of propagation associated to the perturbative degree of freedom.
This causal cone is inherently different for different helicities
---~scalar, vector, or tensor~--- and its structure determines
whether the modes are healthy or not.

Furthermore, matter is assumed to couple universally to a single
metric in order to pass stringent fifth-force experiments. This
introduces the matter causal cone, in addition to gravity cones,
and the physical metric, associated to geodesic free-fall, which
matter couples to. The prototype example is Brans-Dicke (BD)
scalar-tensor theory (see for example \cite{Sotiriou:2015lxa}).
There, the Jordan frame is the physical frame, whose associated
metric gives geodesic free fall. The more calculation-friendly
Einstein frame is related to the physical frame \textit{via} a
specific conformal transformation involving the metric and the
scalar degree of freedom.
Note that for Horndeski theory and beyond, modes of different
species (in our example scalar and tensor) mix together and it is
in general only for the most symmetric backgrounds that one
manages to demix them.

The effect of multiple causal cones and mixings is that species
---~such as gravitons~--- can now have subluminal or superluminal
propagation with respect to photons (matter light cone).
However, these multiple possibilities have been recently
constrained by a single observation. The simultaneous detection
of gravity wave event GW170817 \cite{TheLIGOScientific:2017qsa}
and light emanating from the same source \cite{Monitor:2017mdv}
at 40~Mpc distance strongly restricts the graviton (spin 2)
causal cone and that of light to be essentially identical, just
like in GR. This restricts the variety of modified gravity
theories to a subclass of theories with gravitons propagating at
the speed of light, i.e., such that $c_\text{g}=1$ in the
physical frame~\cite{Ezquiaga:2017ekz,
Creminelli:2017sry,Sakstein:2017xjx,Lombriser:2015sxa,Lombriser:%
2016yzn,Bartolo:2017ibw,Baker:2017hug}.
On the other hand, in some cases, it is technically easier to
work in a non-physical frame, where the metric is disformally
related to the physical metric.
For Horndeski theory, for example, one can find a disformal
transformation of the metric that brings it to a unit propagation
speed \cite{Ezquiaga:2017ekz} ($c_\text{g}=1$) theory of
EST/DHOST type \cite{Crisostomi:2016czh,Achour:2016rkg}. Note
that this is not possible for any Horndeski theory; for example,
$G_5$ interactions involving the Gauss-Bonnet term are excluded
\cite{Ezquiaga:2017ekz, Creminelli:2017sry}\footnote{By
excluded, throughout this paper, we refer to the cases where the
extra mode is a dark energy field, giving an effective
acceleration to the universe at late times. If the extra mode,
say a scalar, is not varying at cosmological scales, but only
locally, it may not influence gravitational waves in their 40~Mpc
journey to Earth detectors.}. In a recent paper
\cite{Babichev:2017lmw}, considering a vacuum black hole
background, we showed how the right choice of disformal
transformation can ensure that the graviton speed is the same as
that of light.
Although we started from a shift-symmetric $G_4$ Horndeski
theory, which is excluded in the physical frame, there exists a
specific disformal transformation \cite{Ezquiaga:2017ekz} upon
which the graviton causal cone is identical to that of light as
demanded by observation. The physical metric is a disformed
metric of the Horndeski action, which mixes scalar and metric
perturbations and brings us to a strictly $c_\text{g}=1$ theory.
In this case the initial Horndeski theory plays the role of the
Einstein frame, whereas the target EST/DHOST theory
\cite{Crisostomi:2016czh,Achour:2016rkg} plays the role of the
Jordan frame, in analogy to the familiar BD theory cited above.

In this paper we will see how, starting from the causal cone
structure of propagating degrees of freedom, one can infer if the
modes in question are healthy modes ---~in other words that they
are not modes generating ghost or gradient instabilities.
In particular, the sign of the determinant of the effective
metric defines the hyperbolicity condition, which if satisfied,
means that a particular solution is safe from imaginary speeds of
propagation and therefore gradient instabilities.
On the other hand, the local orientation of the cone tells us
about absence/presence of ghost modes.
Both requirements allow the local definition of a causal cone of
propagation which then guarantees a healthy associated mode.

A complementary way to find the good or sick nature of
propagating modes is often described \textit{via} the associated
Hamiltonian density of the modes in question. Once the effective
action for the mode is known, one defines the conjugate momentum
and writes down the Hamiltonian density of the associated field.
It is known that if the Hamiltonian density is bounded from
below, then the ground state is of finite energy and necessarily
stable. The contrary is often assumed to be true: If a
Hamiltonian is unbounded from below, then the system is unstable
and admits ghost or gradient instabilities.

One of the main aims of our paper is to explicitly show that the
above inverse statement is not always true. In other words, if a
Hamiltonian density is unbounded from below, this does not
necessarily signify that the mode in question generates a ghost
or gradient instability. The reasoning is simple although it goes
against standard lore originating from particle physics or highly
symmetric backgrounds associated to
Friedmann-Lema\^{\i}tre-Robertson-Walker (FLRW) cosmology. The
Hamiltonian is not a scalar quantity and therefore depends on the
coordinate system it is associated with. As such we will
explicitly see that Hamiltonian densities can be unbounded by
below but under a coordinate transformation can be transformed to
a bounded density. The key point will be the coordinate system on
which the Hamiltonian is to be defined in relation to the
effective causal cones.

In fact, we will see that the coordinate system will have to be
of a certain ``good'' type in order for the Hamiltonian density
to be conclusive. For our purposes, we will restrict ourselves to
configurations where essentially the problem is mathematically
2-dimensional and involves the definition of a good timelike and
spacelike direction. This includes the case for planar,
cylindrical or spherical symmetry, for example. A ``good''
coordinate system will involve the existence of a common timelike
direction for all causal cones. Secondly it will involve the
existence of a common spacelike direction exterior to all causal
cones.\footnote{A causal cone represents an open set whose
interior is bounded by the characteristics of the cone. The
complementary of this set with boundary is an open set which is
the exterior of the cone.} If such a coordinate system exists,
then we will show that the Hamiltonian density is bounded from
below and the system is stable. If such a coordinate system does
not exist, on the contrary, then the Hamiltonian density is
always unbounded from below. The relevant criteria emerging from
the causal cones will inevitably lead to the knowledge of ghost
or gradient instabilities present in the system.

We will explicitly show all this for a general situation with two
propagating degrees of freedom with different causal structures
in Sec.~\ref{Sec2},
and apply it to the known stability criteria of k-essence. We
will see explicitly how stability criteria are satisfied for
well-defined causal cones and how on the contrary, the
unboundedness of Hamiltonian densities can lead to wrong
conclusions if not associated to a ``good'' coordinate system. We
will then move on, in Sec.~\ref{sec:Horndeski}, to apply our
causal cone criteria to investigate the stability of a specific
Horndeski theory admitting a non-trivial background black-hole
solution \cite{Babichev:2013cya}. The family of strongly
gravitating solutions admit a time-dependent scalar field which
is asymptotically a dark energy field allowing de Sitter
acceleration inherently different from the vacuum cosmological
constant.
The mixed combination of space and time dependence for the
scalar, as well as the higher order nature of the theory, leads
to causal scalar and tensor cones which are quite complex. This
is the reason why the Hamiltonian analysis of
\cite{Ogawa:2015pea} gave the wrong
conclusion\footnote{References
\cite{Takahashi:2015pad,Takahashi:2016dnv,Kase:2018voo} used
similar arguments, and accordingly obtain too restrictive
conditions for stability. Reference \cite{Maselli:2016gxk} also
uses these arguments, but it proves the \textit{stability} of the
odd-parity modes outside neutron stars, and this is correct.},
stating the generic instability of such black holes for any
coupling constants of the theory.
Although the Hamiltonian associated with the graviton is
unbounded by below in Schwarzschild coordinates, we will see that
it is bounded by below in an appropriate coordinate system. The
graviton and matter causal cones indeed keep compatible
orientations, and can actually be chosen to exactly coincide at
any spacetime point in order to satisfy the gravity speed
constraint imposed by the GW170817 event. We will additionally
complete this analysis by deriving the scalar causal cone, and
showing that it also has a compatible orientation with the two
previous cones for a certain range of parameters of the model. We
compute this last cone by studying the $\ell=0$ perturbations.
We will conclude in Sec.~\ref{sec:conc}.

\section{Causal cones and Hamiltonian in a general coordinate system}
\label{Sec2}

Standard theories minimally coupling all fields to one metric
tensor $g_{\mu\nu}$ possess a single causal cone, defined by
$ds^2 \equiv g_{\mu\nu} dx^\mu dx^\nu = 0$, or equivalently by
$g^{\mu\nu} k_\mu k_\nu = 0$ for a wave vector $k_\mu$,
$g^{\mu\nu}$ denoting as usual the inverse of $g_{\mu\nu}$.
This is no longer the case when at least two fields are coupled
to different metric tensors which are not proportional
---~even though they are generally related to each others.
For instance, all Galileon models [aside from the simplest
$(\partial_\mu\varphi)^2$ Lagrangian for a scalar field
$\varphi$] and their generalizations called beyond-Horndeski
theories \cite{Horndeski:1974wa,Fairlie,Nicolis:2008in,%
Deffayet:2009wt,Deffayet:2009mn,deRham:2010eu,Deffayet:2011gz,%
Deffayet:2013lga,Bettoni:2013diz,Zumalacarregui:2013pma,%
Gleyzes:2014dya,Lin:2014jga,Gleyzes:2014qga,Deffayet:2015qwa,%
Langlois:2015cwa,Langlois:2015skt,Crisostomi:2016tcp,%
Crisostomi:2016czh,Achour:2016rkg,deRham:2016wji,%
BenAchour:2016fzp}, predict that the spin-2 and spin-0 degrees of
freedom propagate in different effective metrics, which depend on
the background solution. And these two effective metrics actually
also generically differ from $g_{\mu\nu}$, to which one may (or
may not) choose that matter fields are universally coupled.

The simplest example is k-essence
\cite{Bekenstein:1984tv,ArmendarizPicon:1999rj,%
Chiba:1999ka,ArmendarizPicon:2000ah},
i.e., a Lagrangian for the scalar field given by a
\textit{non-linear} function $f$ of the standard kinetic term,
$\mathcal{L} = -\frac{1}{4}f(X)$, where
$X\equiv g^{\mu\nu}\partial_\mu\varphi\partial_\nu\varphi$. If
one writes the scalar field as $\varphi = \bar\varphi+\chi$,
where $\bar\varphi$ denotes the background solution and $\chi$ a
small perturbation, one finds that the second-order expansion of
this Lagrangian reads $\mathcal{L}_2 =-\frac{1}{2}
\mathcal{S}^{\mu\nu} \partial_\mu\chi \partial_\nu\chi$, where
\begin{equation}
\mathcal{S}^{\mu\nu} = f'(\bar X) g^{\mu\nu} + 2 f''(\bar X)
\nabla^\mu\bar\varphi \nabla^\nu\bar\varphi,
\label{Eq:GmunuKessence}
\end{equation}
$f'$ and $f''$ being the first and second derivatives
of function $f$ with respect to its argument $\bar X$
\cite{ArmendarizPicon:1999rj,Babichev:2006vx,Bruneton:2006gf,%
Bruneton:2007si,Babichev:2007dw}. This means that the spin-0
degree of freedom $\chi$ propagates in an effective
metric\footnote{We use the notation $\mathcal{S}^{\mu\nu}$
for the effective metric in which the $\mathcal{S}$calar degree of
freedom propagates, while $\mathcal{G}^{\mu\nu}$ will be used
in Sec.~\ref{sec:effmetrics} to denote the effective metric in
which spin-2 degrees of freedom ($\mathcal{G}$ravitons) propagate.}
$\mathcal{S}^{\mu\nu}$, which is not proportional to $g^{\mu\nu}$
as soon as $f''(\bar X) \neq 0$ and the background solution has a
non-vanishing gradient $\partial_\mu\bar\varphi$, and they define
thus different causal cones. In this simple case, one can show
that the spin-2 degrees of freedom (the gravitons, perturbations
of the metric tensor) do propagate in the initial metric
$g^{\mu\nu}$. To simplify this example even further, one may
actually consider it in \textit{flat} spacetime, i.e., without
any graviton, while universally coupling matter to $g_{\mu\nu}$.
Then we still have at least two fields (matter and the spin-0
degree of freedom $\chi$) which propagate in different metrics,
defining two different causal cones.

\begin{figure}
\includegraphics[width=\textwidth]{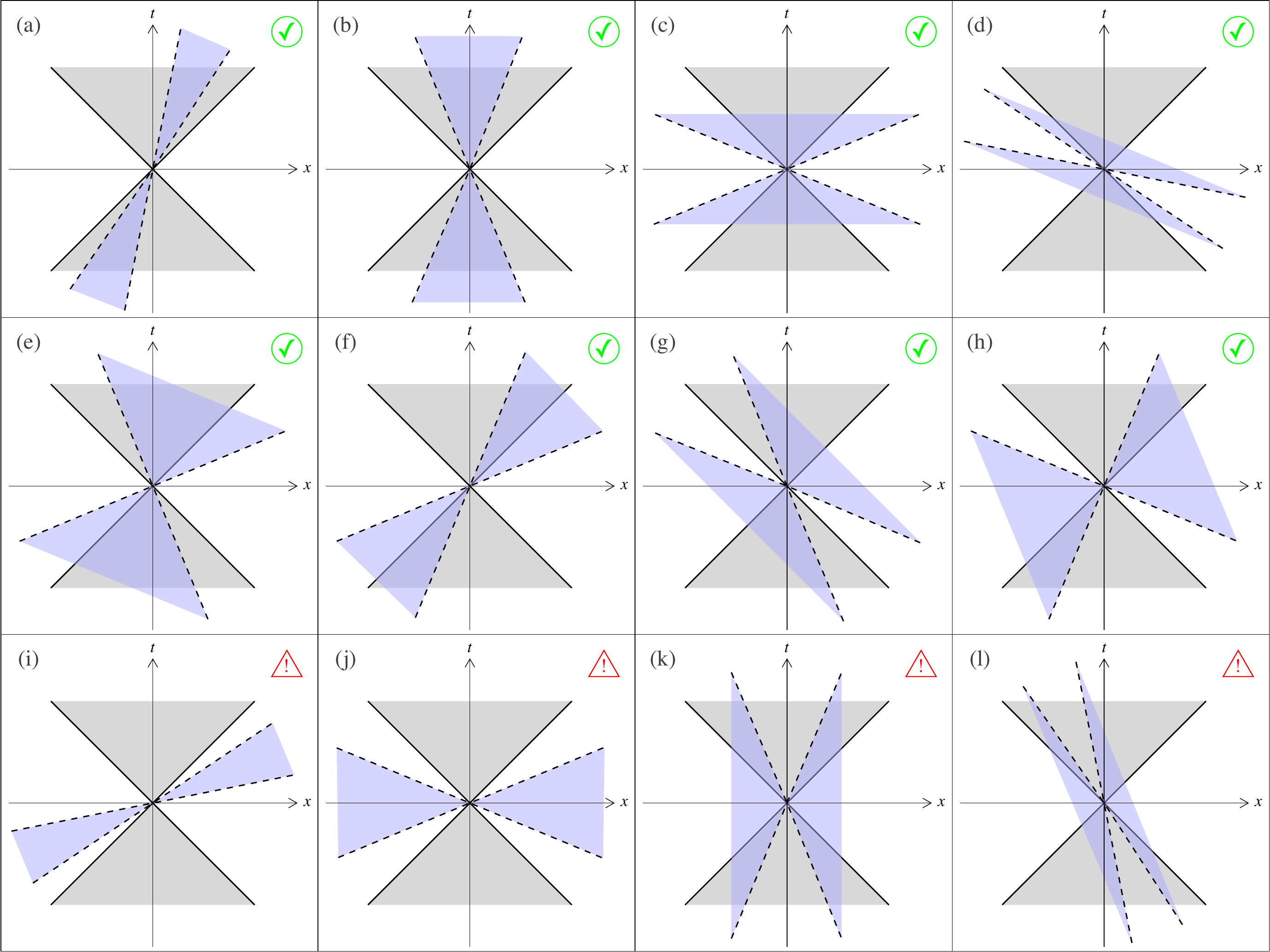}
\caption{Possible relative orientations of two causal cones,
in a coordinate system such that the grey cone with solid lines
appears at $\pm45^\circ$. We do not plot the equivalent
configurations exchanging left and right, and do not consider
the limiting cases where some characteristics coincide. The
first row (a)--(d) are safe cases in which the two metrics
can be diagonalized simultaneously by an appropriate choice
of coordinates ---~corresponding then to panels (b) or (c).
Although the kinetic contribution to their Hamiltonian density is
unbounded by below in cases (a) and (d), it is positive in (b)
and (c). The second row (e)--(h) are again safe cases, for which
the kinetic contribution to the Hamiltonian density can be proven
to be positive in an appropriate coordinate system, actually
corresponding to case (e), but the two metrics cannot be
simultaneously diagonalized\protect\footnote{Let us recall that
two quadratic forms can always be simultaneously diagonalized if
at least one of them is positive (or negative) definite. Here
both of our metrics have hyperbolic signature, and this is the
reason why the non-simultaneously diagonalizable cases (e)--(h)
are possible.}. The third row (i)--(l) are unstable cases,
for which the two metrics can be simultaneously diagonalized as
in (j) and (k), but they have then opposite signatures in this
$(t,x)$ subspace. Their total Hamiltonian density remains
unbounded by below in all coordinate systems.}
\label{fig:CausalCones}
\end{figure}

The conditions for such a k-essence theory to be stable and have
a well-posed Cauchy problem have been written several times in
the literature \cite{Aharonov:1969vu,ArmendarizPicon:1999rj,%
Babichev:2006vx,Bruneton:2006gf,Bruneton:2007si,Babichev:2007dw},
and we shall rederive them at the end of the present Section from
our general analysis. They read\footnote{We choose the
mostly-plus signature convention for the metric $g_{\mu\nu}$.}
$f'(\bar X) > 0$ and $2 \bar X f''(\bar X) + f'(\bar X) > 0$.
When the background scalar gradient $\partial_\mu\bar\varphi$ is
timelike with respect to $g^{\mu\nu}$, the causal cones can be
represented as panels (a), (b), (c) or (d) of
Fig.~\ref{fig:CausalCones}, where the grey cone (with solid
lines) is defined by $g^{\mu\nu}$ and the blue one (with dashed
lines) by $\mathcal{S}^{\mu\nu}$. Panel (a) is actually
transformed into (b), and (d) into (c), if one chooses
a coordinate system such that the spatial gradients
$\partial_i\bar\varphi$ vanish, the vector
$\partial_\mu\bar\varphi$ pointing then exactly
in the time direction.

Panels (a) and (b) correspond to $f''(\bar X) < 0$, and mean that
the spin-0 degree of freedom $\chi$ propagates slower than light
(which is the fastest matter field). Panels (c) and (d)
correspond to $f''(\bar X) > 0$, and describe a superluminal
scalar field, but this does not lead to any causality problem as
soon as this dashed cone remains always a cone, with a non-empty
exterior where one may define Cauchy surfaces to specify initial
data. This has already been discussed in detail in the literature
\cite{Aharonov:1969vu,Babichev:2006vx,%
Bruneton:2006gf,Bruneton:2007si,Babichev:2007dw}.
Paradoxes only occur when one wants to specify initial data on
the $t=0$ hypersurface in the situation of panel (d): This is
forbidden because this hypersurface is not spacelike with respect
to the dashed cone. Note that panels (b) and (c) are actually
equivalent if one exchanges the meaning of the colors. If one
chooses a coordinate system such that the scalar causal cone is
at $\pm45^\circ$ (grey cone with solid lines), then matter
propagates within the dashed blue cone, and the case of a
superluminal scalar perturbation $\chi$ now corresponds to panels
(a)--(b). Causality becomes then more obvious than in panel (d).

Independently of the specific form (\ref{Eq:GmunuKessence}) taken
by the effective metric $\mathcal{S}^{\mu\nu}$ in the case of
k-essence, let us now consider any possible
$\mathcal{S}^{\mu\nu}$ in which a field $\chi$ propagates, to
discuss all the cases of Fig.~\ref{fig:CausalCones}. To simplify,
we shall assume that the standard metric $g_{\mu\nu}$ (to which
matter is assumed to be universally coupled) is flat. If the
Lagrangian defining the dynamics of $\chi$ reads as before
$\mathcal{L}_2 =-\frac{1}{2}\mathcal{S}^{\mu\nu} \partial_\mu\chi
\partial_\nu\chi$, where we focus only on the kinetic term, then
the conjugate momentum is defined as
\begin{equation}
p \equiv \frac{\partial \mathcal{L}_2}{\partial \dot\chi}
=-\mathcal{S}^{00}\dot\chi - \mathcal{S}^{0i}\partial_i\chi,
\label{Eq:p}
\end{equation}
and the contribution of this field $\chi$ to the Hamiltonian
density reads
\begin{equation}
\mathcal{H}_2 = p\,\dot\chi - \mathcal{L}_2 =
-\,\frac{1}{2\mathcal{S}^{00}}
\left(p+\mathcal{S}^{0i}\partial_i\chi\right)^2
+\frac{1}{2}\,\mathcal{S}^{ij} \partial_i\chi \partial_j\chi.
\label{Eq:H2}
\end{equation}
Note that its positiveness depends only on $\mathcal{S}^{00}$
and $\mathcal{S}^{ij}$, but not on the mixed components
$\mathcal{S}^{0i}$, although we shall see that they are actually
crucial for the stability analysis. Stability is indeed a
physical (observable) statement, which should be coordinate
independent, whereas the Hamiltonian density is \textit{not}
a scalar and depends thus on the coordinate system.

To simplify even further the discussion, let us assume that
$\mathcal{S}^{\mu\nu}$ is of the form
\begin{equation}
\begin{pmatrix}
\mathcal{S}^{00} & \mathcal{S}^{01} & 0 & 0 \\
\mathcal{S}^{01} & \mathcal{S}^{11} & 0 & 0 \\
0 & 0 & \mathcal{S}^{22} & 0 \\
0 & 0 & 0 & \mathcal{S}^{33}
\label{Eq:matrixG}
\end{pmatrix},
\end{equation}
with $\mathcal{S}^{22}\geq 0$ and $\mathcal{S}^{33}\geq 0$, and
let us focus on the $(t,x)$ subspace as in
Fig.~\ref{fig:CausalCones}. [In the neighborhood of a spherical
body, for instance, it is natural to choose spherical coordinates
where $\mathcal{S}^{\theta\theta} = 1/r^2$ and
$\mathcal{S}^{\phi\phi} = 1/(r^2\sin^2\theta)$ or similar in
generalized Galileon and beyond-Horndeski theories, the
difficulties being restricted to the $(t,r)$ subspace.] In order
for this metric to define a cone, with non-empty interior and
exterior, it is necessary that its determinant be negative:
\begin{equation}
D\equiv \mathcal{S}^{00}\mathcal{S}^{11}
- \left(\mathcal{S}^{01}\right)^2 < 0.
\label{Eq:Det}
\end{equation}
Note that this hyperbolicity condition does depend on the
off-diagonal component $\mathcal{S}^{01}$, contrary to the sign
of Hamiltonian (\ref{Eq:H2}) above. The \textit{inverse}
$\mathcal{S}^{-1}_{\mu\nu}$ of matrix (\ref{Eq:matrixG}) [its
exponent $-1$ being explicitly written in order not to confuse it
with $g_{\mu\lambda}g_{\nu\rho}\mathcal{S}^{\lambda\rho}$] reads
in the $(t,x)$ subspace
\begin{equation}
\begin{pmatrix}
\mathcal{S}^{11} & -\mathcal{S}^{01} \\
-\mathcal{S}^{01} & \mathcal{S}^{00}
\end{pmatrix}/D.
\label{Eq:Ginverse}
\end{equation}
We can thus conclude that when $\mathcal{S}^{\mu\nu}$ indeed
defines a cone, then $\mathcal{S}^{-1}_{00}$ has the opposite
sign of $\mathcal{S}^{11}$, and $\mathcal{S}^{-1}_{11}$ the
opposite sign of $\mathcal{S}^{00}$.

Let us now consider the various cone orientations of
Fig.~\ref{fig:CausalCones}. In the situation of panel (a), the
time axis is outside the dashed (blue) cone defined by
$\mathcal{S}^{\mu\nu}$. This means that $\mathcal{S}^{-1}_{00}\,
dt\, dt > 0$, and therefore $\mathcal{S}^{11} < 0$. This implies
that the Hamiltonian density (\ref{Eq:H2}) is unbounded by below
because of the contribution of $\frac{1}{2}\,\mathcal{S}^{ij}
\partial_i\chi \partial_j\chi$, when $\partial_1\chi$ is large
enough (and $p$ is chosen to compensate
$\mathcal{S}^{0i}\partial_i\chi$). This conclusion remains the
same for all panels of this Figure in which the time axis is
outside the dashed cone, namely (f), (h), (i), (j), and (k). On
the contrary, when the time axis is within the dashed cone (in
all other panels of Fig.~\ref{fig:CausalCones}), this corresponds
to $\mathcal{S}^{11} > 0$, and the second term of Hamiltonian
(\ref{Eq:H2}) is thus positive.

Similarly, in the situation of panel (d), the $x$ axis is within
the dashed cone, therefore $\mathcal{S}^{-1}_{11}\, dx\, dx < 0$,
which implies $\mathcal{S}^{00} > 0$. In this case, the
Hamiltonian density (\ref{Eq:H2}) is unbounded by below because
of the contribution of its first term
$-\left(p+\mathcal{S}^{0i}\partial_i\chi\right)^2/
\left(2\mathcal{S}^{00}\right)$. This conclusion remains the same
for all panels in which the $x$ axis is inside the dashed cone,
namely (g), (h), (j), (k), and (l). In all other panels, the $x$
axis is outside the dashed cone, therefore $\mathcal{S}^{00} < 0$
and the first term of Hamiltonian (\ref{Eq:H2}) is thus positive.

Note that panels (h), (j) and (k) have both their time axis
outside the dashed cone and their $x$ axis within it. This means
that the Hamiltonian density (\ref{Eq:H2}) is always negative,
while that corresponding to matter (coupled to $g_{\mu\nu}$ and
propagating thereby in the solid grey cone) is always positive.
It thus naively seems that any coupling between matter and
$\chi$, or any indirect coupling \textit{via} another field (for
instance gravity), will lead to deadly instabilities. This is
indeed the case for panels (j) and (k), but not for panel (h).
Indeed, if one chooses another coordinate system such that the
new time $t'$ lies within the intersection of both cones
(superposition of the grey and blue regions), and the new spatial
direction $x'$ is outside both cones (white region), then one
gets simultaneously the four conditions $g'^{00} < 0$,
$g'^{11} > 0$, $\mathcal{S}'^{00} < 0$, and
$\mathcal{S}'^{11} > 0$. Therefore, both the Hamiltonian density
(\ref{Eq:H2}) for the spin-0 degree of freedom $\chi$ and its
analogue for matter are positive in this coordinate system.
This suffices to prove that no instability can be caused by the
kinetic terms in the situation of panel (h).

Let us recall that when a total Hamiltonian density (including
all interacting fields) is bounded by below, then the
lowest-energy state is necessarily stable. It is indeed
impossible to reach a higher energy state (for any field) without
violating energy conservation. But note that the converse theorem
does not exist, as underlined by the reasoning above: A
Hamiltonian density which is unbounded by below does not always
imply an instability. In panel (h) of Fig.~\ref{fig:CausalCones},
this Hamiltonian was the sum of the positive contribution of
matter and of the (always) negative contribution of the spin-0
field $\chi$, but we saw that there exist other coordinate
systems in which both contributions are simultaneously positive.

To understand this better, let us just consider the boosts of
special relativity in flat spacetime, instead of the arbitrary
coordinate transformations allowed in GR. Then the metric
$g^{\mu\nu}$ in which matter propagates always reads
$\text{diag}(-1,1,1,1)$, and it defines the solid (grey) cones of
Fig.~\ref{fig:CausalCones}. In the simple cases of panels (b) and
(c), the components of $\mathcal{S}^{\mu\nu}$ in the $(t,x)$
subspace read $k^2\text{diag}(-1/c_\text{s}^2,1)$, where $k$
is a non-vanishing constant and $c_\text{s}$ is the velocity
corresponding to the characteristics of the dashed (blue)
cone\footnote{We set $c=1$ for the velocity of light,
corresponding to the solid (grey) cone.}. Indeed, the wave
equation for the spin-0 field $\chi$ reads $\mathcal{S}^{\mu\nu}
\partial_\mu\partial_\nu\chi = 0$, and it admits as solutions
arbitrary functions of $(x\pm c_\text{s} t)$. Panel (b)
corresponds to $c_\text{s}^2 < 1$ while panel (c) to
$c_\text{s}^2 > 1$. If we now perform a boost of velocity $-v$,
we find that the components of $\mathcal{S}'^{\mu\nu}$ in the
new coordinate system read
\begin{equation}
\frac{k^2}{c_\text{s}^2(1-v^2)}
\begin{pmatrix}
-1 + v^2 c_\text{s}^2\quad & v (1-c_\text{s}^2) \\
v (1-c_\text{s}^2)\quad & c_\text{s}^2-v^2
\end{pmatrix}.
\label{Eq:BoostedG}
\end{equation}
We thus immediately see that $\mathcal{S}'^{11} < 0$
(with $\mathcal{S}'^{00}$ still negative) when we choose
$|c_\text{s}|<|v|<1$ in the case of panel (b), i.e., that
we obtain the situation of panel (a), as described below
Eq.~(\ref{Eq:Ginverse}). Although we started from the stable
situation of panel (b), in which the total Hamiltonian density is
positive, we thus find that the contribution of the spin-0 degree
of freedom is no longer bounded by below in this boosted frame
corresponding to panel (a). This is the main lesson: The
unboundedness by below of the Hamiltonian density is a mere
coordinate effect in the present situation, and it has no
physical meaning. The model is stable, but one is not computing
the ``right'' quantity in the boosted frame of panel (a). [We
shall come back to this ``right'' quantity below.]

Note that a negative value of $\mathcal{S}'^{11}$ in the boosted
frame of panel (a) always comes together with a significant
non-zero value of $|\mathcal{S}'^{01}| = |\mathcal{S}'^{10}| >
\sqrt{-D}$, where $D$ is the determinant (\ref{Eq:Det}). The
reason is that this determinant must remain negative in all
coordinate systems ---~and actually remains strictly equal to $D$
when one considers only special-relativistic boosts as here.
These non-zero off-diagonal components of $\mathcal{S}'^{\mu\nu}$
are crucial for the existence of an inverse boost taking us back
to the situation of panel (b), where the total Hamiltonian
density is positive. If they were absent, then the metric
$\text{diag}(\mathcal{S}'^{00}, \mathcal{S}'^{11})$ would be
negative definite, it would not define any causal cone, and the
Cauchy problem would be ill-posed. Note also that the magnitude
of these off-diagonal components of $\mathcal{S}'^{\mu\nu}$ is
also crucial. For instance, panel (i) of
Fig.~\ref{fig:CausalCones} corresponds to $\mathcal{S}'^{00} < 0$
and $\mathcal{S}'^{11} < 0$ like panel (a), and it does satisfy
$|\mathcal{S}'^{01}| > \sqrt{-D}$, but also the inequality
$|\mathcal{S}'^{01}| <
\left|\mathcal{S}'^{00}+\mathcal{S}'^{11}\right|/2$ which leads
to the situation of panel (j) when diagonalizing
$\mathcal{S}^{\mu\nu}$ by an appropriate boost. In this case (j),
the two metrics $g^{\mu\nu}$ and $\mathcal{S}^{\mu\nu}$ have
opposite signatures in the $(t,x)$ subspace, so that the spin-0
degree of freedom $\chi$ behaves as a ghost in this subspace, and
the model is unstable as soon as $\chi$ is somehow coupled to
matter (including indirectly, e.g., \textit{via} gravity).

Let us now apply a boost to the case of panel (c) of
Fig.~\ref{fig:CausalCones}. If we choose $|c_\text{s}|^{-1} < |v|
< 1$, then we find from Eq.~(\ref{Eq:BoostedG}) that
$\mathcal{S}'^{00} > 0$ (with $\mathcal{S}'^{11}$ still
positive), i.e., we obtain the situation of panel (d). Here
again, as described above, we thus find that the contribution of
the spin-0 degree of freedom to the Hamiltonian density is no
longer bounded by below in this boosted frame, whereas is was
positive in the initial frame corresponding to panel (c). [The
fact that the first term of (\ref{Eq:H2}), proportional to
$\dot\chi^2$, becomes negative is related to the wrong
time-orientation\footnote{On the other hand, the possible
negative values of Hamiltonian (\ref{Eq:H2}) in the previous case
of panel (a) is less obvious, since the null vectors $N^\mu$
(with respect to $\mathcal{S}'^{-1}_{\mu\nu}$) always remain
future-oriented. In that case, negative values are caused by the
second term of (\ref{Eq:H2}) involving the spatial derivative
$\partial_1\chi$, and they are thus caused by a specific spatial
dependence of the initial data.} of the null vector $N^\mu$ with
respect to $\mathcal{S}'^{-1}_{\mu\nu}$ in the boosted frame of
panel (d): When it points towards positive values of $x'$, it
seems to go backwards with respect to time $t'$. As underlined
above, the hypersurface $t'=0$ cannot be consistently used to
specify initial data in this case, since it is not spacelike with
respect to $\mathcal{S}'^{-1}_{\mu\nu}$, therefore the sign of
Hamiltonian (\ref{Eq:H2}) at $t'=0$ does not have much meaning
anyway.] The conclusion is the same as before: The unbounded by
below Hamiltonian in the boosted frame of panel (d) is a mere
coordinate effect, without any physical meaning, and the model is
actually stable, as proven by the positive total Hamiltonian
density in the frame of panel (c).

It is also instructive to compute the energy of a system in a
boosted frame (still in flat spacetime, to simplify the
discussion). Although it differs from $g^{\mu\nu}$, the effective
metric $\mathcal{S}^{\mu\nu}$ is a tensor; see for instance
Eq.~(\ref{Eq:GmunuKessence}) for the particular case of
k-essence. Therefore, the Lagrangian $\mathcal{L}_2
=-\frac{1}{2}\mathcal{S}^{\mu\nu} \partial_\mu\chi
\partial_\nu\chi$ is diffeomorphism invariant, and this
implies that four Noether currents are conserved. They read
\begin{equation}
-T_\mu^\nu \equiv
\frac{\delta \mathcal{L}_2}{\delta (\partial_\nu \chi)}\,
\partial_\mu \chi - \delta_\mu^\nu\, \mathcal{L}_2,
\label{eq:Noether}
\end{equation}
where the index $\mu$ specifies which of the four currents is
considered, $\nu$ denotes its components, and $\delta_\mu^\nu$ is
the Kronecker symbol. [A global minus sign is introduced in
definition (\ref{eq:Noether}) so that the mixed component $T^0_0$
denotes the opposite of the energy density, like in general
relativity.] The current conservation reads as usual
$\partial_\nu T^\nu_\mu = 0 \Leftrightarrow \partial_0 T^0_\mu +
\partial_i T^i_\mu = 0$. When integrating this identity over a
large spatial volume $V$ containing the whole physical system
under consideration, the spatial derivatives become vanishing
boundary terms, and one gets the standard conservation laws for
total energy and momentum, $\partial_t P_\mu = 0$, with $P_\mu
\equiv -\int\!\!\!\int\!\!\!\int_V T_\mu^0\, d^3 x$. For $\mu =
0$, the energy density $-T_0^0$ coincides with the on-shell value
of the Hamiltonian density (\ref{Eq:H2}). As recalled above, if
it is bounded by below, then the lowest-energy state must be
stable. But it should be underlined that the three components of
the total momentum $P_i$ are also conserved, and that the
components $-T_i^0 = p\,\partial_i\chi$ [with $p$ still given by
Eq.~(\ref{Eq:p})] have no preferred sign, since there is no
privileged spatial direction. When changing coordinates, the
total 4-momentum of the system becomes $P'_\lambda = (\partial
x^\mu/\partial x'^\lambda)P_\mu$, and in particular, the energy
gets mixed with the initial 3-momentum, $P'_0 = (\partial
x^\mu/\partial x'^0)P_\mu$, or simply $P'_0 = (P_0+ v
P_1)/\sqrt{1-v^2}$ for a mere boost of velocity $v$ in the $x$
direction. Of course, $g^{\mu\nu}P_\mu P_\nu$ (as well as
$\mathcal{S}^{\mu\nu}P_\mu P_\nu$) is a scalar quantity, and it
remains thus invariant under coordinate transformations. However,
it is not always negative, contrary to the standard ``minus rest
mass squared'' in special relativity, therefore the magnitude of
the spatial components $P_i$ is not always bounded by $P_0$. For
instance, in panels (c) or (d) of Fig.~\ref{fig:CausalCones}, a
scalar field perturbation propagating outside the solid (grey)
cone obviously corresponds to a positive $g^{\mu\nu}P_\mu P_\nu$,
i.e., a spacelike $P_\mu$ with respect to $g^{\mu\nu}$. It is
thus clear that a negative value of $P'_0 = (P_0+ v P_1)/
\sqrt{1-v^2}$ is reachable for a large enough boost velocity $|v|
< 1$. The fact that $P'_0$ can also become negative in the case
of panel (a) is much less obvious, but it can be checked that it
coincides with the spatial integral of the on-shell expression of
Hamiltonian (\ref{Eq:H2}) with the boosted effective metric
(\ref{Eq:BoostedG}). In such a case, a large enough boost
velocity $|c_\text{s}|<|v|<1$ generates a negative
$\mathcal{S}'^{11}$, and thereby a possibly negative Hamiltonian
(\ref{Eq:H2}), when initial data on the $t'=0$ hypersurface are
chosen with a large spatial gradient $\partial_{x'}\chi$ (but a
small $\partial_{t'}\chi$). Up to now, we are merely rephrasing
our previous conclusions with a slightly different viewpoint. But
what is more interesting is to understand why situations like
panels (a) or (d) of Fig.~\ref{fig:CausalCones} are stable in
spite of their Hamiltonian density (\ref{Eq:H2}) which is
unbounded by below. The reason is simply that not only their
total energy $P'_0$ is conserved, but also their 3-momentum
$P'_i$. And it happens that the linear combination $(P'_0- v
P'_1)/\sqrt{1-v^2}$, which is thus also conserved, is bounded by
below, since it obviously gives the positive expression of $P_0$
in the initial frame of panels (b) or (c). In other words,
stability is not ensured by the boundedness by below of the
Hamiltonian density, in the present case, but by that of the
linear combination $-T'^0_0 + v T'^0_1$. In more general
situations involving arbitrary coordinate transformations, the
initial energy $P_0$ which is bounded by below is again a linear
combination of conserved quantities in the new frame, $P_0 =
(\partial x'^\mu/\partial x^0)P'_\mu$.

In conclusion, although the Hamiltonian density is not bounded by
below in the situations corresponding to panels (a), (d), (f),
(g) and (h) of Fig.~\ref{fig:CausalCones}, there exists a choice
of coordinates mapping them to panels (b), (c) or (e), where the
new total Hamiltonian density is bounded by below. This suffices
to guarantee the stability of the lowest-energy state, as
computed in this new coordinate system. The only generically
unstable cases correspond to the third row of
Fig.~\ref{fig:CausalCones}, panels (i) to (l), because their
total Hamiltonian density is never bounded by below in any
coordinate system. They are such that the matrix
$\mathcal{S}^{\mu\lambda} g_{\lambda\nu}$ is diagonalizable and
possesses two negative eigenvalues. Conversely, it is easy to
write the inequalities needed on the components of the effective
metric $\mathcal{S}^{\mu\nu}$ to be in the eight safe cases
corresponding to the first two rows, panels (a) to (h): In
addition to the hyperbolicity condition (\ref{Eq:Det}), one
just needs
\begin{equation}
\mathcal{S}^{00} < \mathcal{S}^{11}\quad\text{and/or}\quad
|\mathcal{S}^{00}+\mathcal{S}^{11}| < 2 |\mathcal{S}^{01}|,
\end{equation}
when focusing on the $(t,x)$ subspace in a coordinate system such
that $g_{\mu\nu} = \text{diag}(-1,1)$. But these inequalities are
less enlightening than Fig.~\ref{fig:CausalCones} itself, in
which it is immediate to see whether the two causal cones have
both a common exterior (when one should specify initial data) and
a common interior. When one chooses new coordinates such that
time lies within the cone intersection, and space is outside both
cones, then the total Hamiltonian density caused by kinetic terms
becomes positive.

The above results can also be formulated in a covariant way.
A given solution is stable if and only if \textit{all} effective
metrics (here $g^{\mu\nu}$ and $\mathcal{S}^{\mu\nu}$, but also
$\mathcal{G}^{\mu\nu}$ introduced in Sec.~\ref{sec:Horndeski}
below) are of hyperbolic mostly-plus signature, and it is
possible to find a contravariant vector $U^\mu$ \textit{and}
a covariant vector $u_\mu$ such that
\begin{eqnarray}
g_{\mu\nu} U^\mu U^\nu < 0, &\quad&
\mathcal{S}^{-1}_{\mu\nu} U^\mu U^\nu < 0, \quad \dots,
\label{eq:gUU}\\
g^{\mu\nu} u_\mu u_\nu < 0, &\quad&
\mathcal{S}^{\mu\nu} u_\mu u_\nu < 0, \quad \dots,
\label{eq:guu}
\end{eqnarray}
and
\begin{equation}
T_\mu^\nu U^\mu u_\nu \geq 0,
\label{eq:positiveT00}
\end{equation}
where $T_\mu^\nu$ denotes the Noether currents for all
fields\footnote{Of course, any other \textit{conserved} tensor
constructed from $T_\mu^\nu$ by adding the divergence of an
antisymmetric Belinfante tensor is also allowed
\cite{Belinfante:1939,Bandyopadhyay:1999my}, and in particular
the standard symmetric energy-momentum tensor
$(2/\sqrt{-g})(\delta S_\text{field}/\delta g_{\rho\nu})$ defined
as in general relativity (with one of its indices lowered with
$g_{\mu\rho}$), where $S_\text{field}$ denotes the contribution
of a given field to the action.}, including
Eq.~(\ref{eq:Noether}) for the scalar field.
Equation~(\ref{eq:gUU}) is the covariant way of formulating the
existence of a common interior to all causal cones, where a
``good'' time direction may be chosen, namely $dx^0$ in the
direction of $U^\mu$. Equation~(\ref{eq:guu}) expresses the
existence of a spatial hypersurface exterior to all causal cones,
defined by $u_\mu dx^\mu = 0$, where ``good'' spatial coordinates
may be chosen. Finally, Eq.~(\ref{eq:positiveT00}) states that
the Hamiltonian density is positive in such a ``good'' coordinate
system. Let us underline that $U^\mu$ and $u_\mu$ are generically
\textit{not} related by lowering or raising the index with any of
the effective metrics. This is the crucial difference with
general relativity (with standard minimally coupled fields), in
which a single metric $g_{\mu\nu}$ defines the causal cone of all
degrees of freedom. In this simpler case of GR, the above
conditions boil down to finding a single timelike vector $U^\mu$
for which the usual weak energy condition $T_{\mu\nu} U^\mu U^\nu
\geq 0$ is satisfied. In particular, Eqs.~(\ref{eq:gUU}) and
(\ref{eq:guu}) become then equivalent if one chooses $u_\mu =
g_{\mu\nu} U^\nu$. One may also extend conditions
(\ref{eq:gUU})--(\ref{eq:positiveT00}) by imposing that
\textit{all} ``future-oriented'' contravariant and covariant
vectors $U^\mu$ and $u_\mu$ satisfying (\ref{eq:gUU}) and
(\ref{eq:guu}) respect inequality (\ref{eq:positiveT00}). [By
``future-oriented'', we mean here that these two vectors must
have consistent orientations, i.e., that their scalar product
$U^\mu u_\mu < 0$, otherwise one could change the sign of one of
them without spoiling conditions (\ref{eq:gUU}) nor
(\ref{eq:guu}) but making (\ref{eq:positiveT00}) negative.] This
would be the full generalization of the weak energy condition to
our more subtle case involving several causal cones, and our
previous discussion shows that it would indeed be satisfied if
the solution is stable. However, let us underline again that
stability is actually ensured as soon as one pair of vectors
$U^\mu$ and $u_\mu$ satisfies
Eqs.~(\ref{eq:gUU})--(\ref{eq:positiveT00}).

As an application of the above results, let us rederive the
stability conditions for the effective metric
(\ref{Eq:GmunuKessence}) corresponding to k-essence. Let us
first choose a locally inertial frame such that $g_{\mu\nu} =
\text{diag}(-1,1,1,1)$. Then, if $\partial_\mu \bar\varphi$ is
timelike with respect to $g_{\mu\nu}$, it is always possible to
boost this coordinate system such that $\partial_i\bar\varphi =
0$. We thus get $\mathcal{S}^{\mu\nu} =
\text{diag}\left(\left[-f' + 2\, \dot{\bar\varphi}^2 f''\right],
f',f',f'\right)$. To be in the situation of panels (b) or (c)
of Fig.~\ref{fig:CausalCones}, it is necessary to have
$\mathcal{S}^{00}<0$ and $\mathcal{S}^{xx}>0$, therefore we need
$-f'+ 2\, \dot{\bar\varphi}^2 f'' < 0$ and $f' > 0$. Since $\bar
X = g^{\mu\nu}\partial_\mu\bar\varphi \partial_\nu\bar\varphi =
-\dot{\bar\varphi}^2$ in this specific coordinate system, the
covariant expressions of these conditions are necessarily
$f'(\bar X) > 0$ and $2 \bar X f''(\bar X) + f'(\bar X) > 0$,
as mentioned one paragraph below Eq.~(\ref{Eq:GmunuKessence}).
Note that no condition is imposed on $f''(\bar X)$ alone. The
result remains the same when the background scalar gradient
$\partial_\mu \bar\varphi$ is spacelike (still with respect to
$g_{\mu\nu}$). Then one may choose the $x$ coordinate in its
direction, so that its only non-vanishing component be
$\bar\varphi' \equiv \partial_1\bar\varphi$. In this coordinate
system, the components of the effective metric read
$\mathcal{S}^{\mu\nu} = \text{diag}\left(-f', \left[f' + 2\,
\bar\varphi'^2 f''\right],f',f'\right)$, while $\bar X =
+\bar\varphi'^2$, therefore we recover strictly the same
covariant inequalities. Finally, when $\partial_\mu \bar\varphi$
is a null vector (again with respect to $g_{\mu\nu}$, i.e.,
$\bar X = 0$), it is possible to choose a coordinate system in
which $\partial_\mu \bar\varphi = \left( \dot{\bar\varphi},
\dot{\bar\varphi}, 0,0\right)$, and the non-vanishing components
of the effective metric read $\mathcal{S}^{00} = -f'+ 2\,
\dot{\bar\varphi}^2 f''$, $\mathcal{S}^{11} = f'+ 2\,
\dot{\bar\varphi}^2 f''$, $\mathcal{S}^{01} = \mathcal{S}^{10} =
-2\, \dot{\bar\varphi}^2 f''$, and $\mathcal{S}^{22} =
\mathcal{S}^{33} = f'$. We then find that one of the
characteristics defined by $\mathcal{S}^{\mu\nu}$ coincides with
one of those defined by $g^{\mu\nu}$, corresponding to a velocity
$-1$ for spin-0 perturbations. This is thus a limiting case of
those plotted in Fig.~\ref{fig:CausalCones}. But when $f'(\bar X)
> 0$, consistently with the same covariant inequalities as above,
one finds that the causal cones defined by $g^{\mu\nu}$ and
$\mathcal{S}^{\mu\nu}$ have both a common interior and a common
exterior, and the background solution is thus stable.

\section{Stable black hole solutions in a subclass of (beyond)
Horndeski theories}
\label{sec:Horndeski}

Let us now illustrate our findings with a specific example,
stemming from Horndeski theory. We will discuss certain solutions
of the following action, which has been studied quite a lot due
to its simple self-tuning properties:
\begin{equation}
S_\mathrm{J}[g_{\mu\nu},\varphi] =\displaystyle\int{\sqrt{-g}\,
\mathrm{d}^4x \left[\zeta (R-2\Lambda_\text{bare})+\beta G^{\mu
\nu} \partial_{\mu}\varphi \partial_{\nu}\varphi - \eta \,
\varphi_\lambda^2\right]},
\label{eq:actionJ}
\end{equation}
where we use the simplifying notation $\varphi_\lambda \equiv
\partial_\lambda\varphi$, so that $\varphi_\lambda^2 =
g^{\mu\nu}\partial_\mu\varphi\partial_\nu\varphi$ (which was also
denoted as $X$ in Sec.~\ref{Sec2}).
$\zeta$ is the Planck mass squared divided by $16 \pi$, and
$\eta$, $\beta$ and $\Lambda_\text{bare}$ are some constants.
In terms of standard Horndeski notation, this action corresponds to
\begin{equation}
G_4=\zeta-\frac{\beta}{2} \varphi_\lambda^2 , \qquad G_2=-2\zeta
\Lambda_\text{bare}-\eta \varphi_\lambda^2.
\end{equation}
Static and spherically symmetric black hole solutions of the
above theory were first derived in~\cite{Rinaldi:2012vy} while
they were extended
in~\cite{Babichev:2013cya,Anabalon:2013oea,Minamitsuji:2013ura}
to the case of non-vanishing $\Lambda_\text{bare}$. A new family
of solutions, with a linearly time-dependent scalar field, was
proposed in~\cite{Babichev:2013cya}. These solutions enjoy novel
regularity properties thanks to the time dependence of the
scalar. Some solutions have spacetime metrics that are identical
to their GR counterparts (apart from the value of the
cosmological constant). As a result, they are often referred to
as stealth solutions. More importantly, time dependence of the
scalar field qualifies the scalar to be a dark energy field
responsible for late-time acceleration (as well as self-tuning
properties). These solutions were claimed to be unstable under
linear perturbations~\cite{Ogawa:2015pea}, and more recently the
theory~(\ref{eq:actionJ}) was ruled out observationally. The aim
of the forthcoming section is to show that the former result is
in fact wrong, while the latter crucially depends on how the
metric couples to matter. Put in other words, if the physical
metric to which matter couples minimally is $g_{\mu\nu}$, then
the above theory is ruled out (more precisely, the scalar field
is ruled out as a dark energy candidate). Indeed, the speed of
gravitons in this theory generically deviates from the speed of
light~\cite{Ezquiaga:2017ekz,Creminelli:2017sry} in inconsistency
with the simultaneous observation of gravitational and
electromagnetic waves from the same source,
GW170817~\cite{TheLIGOScientific:2017qsa,Monitor:2017mdv}.
However, it is easy to map the action~(\ref{eq:actionJ}) to a
beyond Horndeski theory in which gravitational waves do travel at
the speed of light in accordance with observations. This has been
checked in weakly curved backgrounds
\cite{Ezquiaga:2017ekz,Creminelli:2017sry} but
also in strongly curved spherically symmetric backgrounds
\cite{Babichev:2017lmw} (and Ref.~\cite{Kase:2018owh} recently
proved so for vector-tensor theories too). To make therefore the
theory~(\ref{eq:actionJ}) viable, the matter action should be
minimally coupled to $\tilde{g}_{\mu \nu}$, the physical metric:
\begin{equation}
\label{eq:disformal}
\tilde{g}_{\mu \nu}=g_{\mu \nu}
-\dfrac{\beta}{\zeta+\dfrac{\beta}{2}
\varphi_\lambda^2}\,\partial_\mu\varphi \partial_\nu \varphi.
\end{equation}
Of course, any metric proportional to this $\tilde{g}_{\mu \nu}$
would also be allowed, since it would not change the causal cone,
even if the conformal factor depends on $\varphi_\lambda^2$.
One should then work with the action
\begin{equation}
S_\mathrm{J}[g_{\mu \nu}, \varphi]+S_\mathrm{m}[\tilde{g}_{\mu
\nu}, \Psi],
\label{eq:goodth}
\end{equation}
where $S_\mathrm{m}$ is some given matter action with matter
fields, collectively denoted as $\Psi$, universally coupled to
the physical metric $\tilde{g}_{\mu \nu}$. In standard
nomenclature for BD gravity, the non-physical $g_{\mu\nu}$ would
be called the ``Einstein frame'' metric. However, its
perturbations do not describe pure spin-2 degrees of freedom in
the present case, because of the kinetic mixing introduced by the
$G^{\mu \nu} \varphi_\mu \varphi_\nu$ term of action
(\ref{eq:actionJ}). We will therefore call $g_{\mu\nu}$ the
``Horndeski frame'' metric rather than the ``Einstein frame''
one. On the other hand, we call $\tilde{g}_{\mu \nu}$ the
``Jordan frame'' physical metric. As in standard BD theory, it is
easier to work in the non-physical frame because the metric
sector is simpler there. We should keep in mind that our analogy
is to be taken with caution, because the frames of the higher
order theories are related disformally~(\ref{eq:disformal}), and
not conformally as in BD theory. Indeed, the disformal
factor~(\ref{eq:disformal}) has been chosen in order to impose a
unit speed for the gravitational waves in the physical (or
Jordan) frame, at least in weakly curved backgrounds. We recently
reported~\cite{Babichev:2017lmw} that the black hole solutions
found in~\cite{Babichev:2013cya} are again black holes with
respect to the physical frame. This is not a trivial result, as a
disformal transformation may change the nature of solutions,
rendering them even singular upon going from one frame to the
other. We will explicitly work out the physical disformed metric
in the next section. Furthermore, we study the stability of some
solutions of the theory~(\ref{eq:goodth}). We do so in the
Horndeski frame, as stability properties carry through upon field
redefinitions~(\ref{eq:disformal}) as long as these are not
singular. A priori, three causal cones must be considered in our
analysis, and must have compatible orientations for the solutions
to be stable: the matter causal cone associated to
$\tilde{g}_{\mu \nu}$, and the cones associated to scalar and
gravitational perturbations (with their associated effective
metrics). Quite remarkably, as we will see, the graviton
perturbation cone will end up being {\it{identical}} to the
matter light cone in the physical frame, demonstrating that
gravity waves travel at same speed as light, even in a strongly
curved region of spacetime (close to the event horizon). This
will effectively reduce the number of causal cones under scrutiny
from three to two. We will see in the next section how to
construct these causal cones and effective metrics in a
spherically symmetric background.

Regarding stability, Appleby and Linder examined
action~(\ref{eq:actionJ}) with vanishing $\Lambda_\text{bare}$ in
a cosmological framework~\cite{Appleby:2011aa}. From the study of
scalar perturbations, they found that there always exists either
a gradient instability or a ghost. This pathology can however be
cured by the introduction of a bare cosmological constant
$\Lambda_\text{bare}$, as we will see below. The stability of the
black hole static solutions was discussed
in~\cite{Kobayashi:2012kh,Kobayashi:2014wsa,Ganguly:2017ort},
based on a more generic theory than~(\ref{eq:actionJ}). The
authors employed the well-established Regge-Wheeler formalism:
they decomposed the perturbations into odd and even modes, each
mode being decoupled of all others at the linear level. Stable
parameter regions were exhibited for the
action~(\ref{eq:actionJ}). Then, Ogawa \textit{et al.} tackled
the case where the scalar field acquires
time-dependence~\cite{Ogawa:2015pea}. They claimed that the
solutions were always unstable, whatever the coupling parameters
of the theory. However, their argument made use of the fact that
the Hamiltonian is unbounded from below; as we argued in the
former section, this cannot be a satisfactory criterion to decide
on the stability of some solution. We show in the last paragraph
of this section that there indeed exist stable black hole
solutions for given parameters. We will first derive the
effective metrics in which graviton and scalar perturbations
respectively propagate.

\subsection{The effective metrics for graviton and scalar
perturbations}
\label{sec:effmetrics}

We will focus our analysis on perturbation theory around
spherically symmetric Schwarzschild-de Sitter solutions. It is
indeed known that the action~(\ref{eq:actionJ}) allows for a
``stealth'' Schwarzschild black hole, as well as Schwarzschild-de
Sitter metrics with a non-trivial scalar profile:
\begin{align}
\label{eq:ansatz}
ds^2 &= -A(r)\, dt^2+\dfrac{dr^2}{B(r)}+ r^2\left(d\theta^2
+\sin^2\theta\, d\phi^2\right),
\\
A(r) &= B(r)=1- \frac{2Gm}{r} - \frac{\Lambda_\text{eff}}{3}\, r^2,
\\
\Lambda_\text{eff} &= -\frac{\eta}{\beta},
\label{EqLambdaEff}
\\
\varphi &= q\left[ t \pm \int\frac{\sqrt{1-A(r)}}{A(r)} dr\right],
\label{Eqphi}
\\
q^2 &= \frac{\eta+ \beta\, \Lambda_\text{bare}}{\eta\,\beta}\,\zeta,
\label{Eqq2}
\end{align}
where $q$ parametrizes the linear time-dependence of the scalar
field, and $m$ corresponds to the mass of the black hole.
The constant $\Lambda_\mathrm{eff}$ plays the role of an
effective cosmological constant, and is a priori independent of
$\Lambda_\text{bare}$ in Eq.~(\ref{eq:actionJ}), with the
velocity integration constant $q$ playing the role of a tuning
integration constant to $\Lambda_\text{bare}$ \textit{via}
relation (\ref{Eqq2}). For consistency, the right-hand side of
Eq.~(\ref{Eqq2}) should be positive; since $\zeta$ is always
positive, we must therefore have
\begin{equation}
(\eta+ \beta\, \Lambda_\text{bare}) \eta \beta>0,
\label{eq:exist1}
\end{equation}
for this solution. A second background solution obtained in the
case $\eta=0$ and $\Lambda_\text{bare}=0$ is the stealth
Schwarzschild black hole solution, which reads
\begin{align}
\label{eq:ansatz2}
ds^2 &= -A(r)\, dt^2 +\frac{dr^2}{B(r)} + r^2\left(d\theta^2
+\sin^2\theta\, d\phi^2\right),
\\
A(r) &= B(r)= 1- \frac{2Gm}{r}
\\
\varphi &= q\left[ t \pm \int\!\frac{\sqrt{1-A(r)}}{A(r)}\, dr\right],
\label{Eqphi2}
\end{align}
and for which $q$ is a free parameter. This solution is
characterized by an asymptotically flat metric, does not have
self tuning properties and is not a limit of the de Sitter black
hole~(\ref{eq:ansatz}).

Before we proceed to the stability analysis let us apply the
disformal transformation~(\ref{eq:disformal}) to the above
background solutions and examine the nature of the physical
metric and scalar field background solution~(\ref{eq:ansatz}).
The family of background solutions verifies the relation
$\varphi_\lambda^2=-q^2$ and this simplifies the disformal
transformation and coordinate transformations thereof. To
simplify notation, we can set
\begin{equation}
\mathcal{F}=-\dfrac{\beta q^2}{\zeta-\frac{\beta}{2}q^2},
\end{equation}
and note that the disformed metric acquires off diagonal terms in
the original $(t,r)$ coordinates due to the $t$ and $r$ scalar
field dependence. We then diagonalize the physical metric using
\begin{equation}
\tilde t=\sqrt{1-\mathcal{F}}\left\{t\mp
\int\!\!\frac{\mathcal{F}\sqrt{1-A(r)}}{A(r)[A(r)-\mathcal{F}]}\,
dr \right\},
\label{Eq:ttilde}
\end{equation}
where the minus sign corresponds to the plus one in
Eqs.~(\ref{Eqphi}) and (\ref{Eqphi2}).
Note that, for this coordinate transformation to be well defined,
one needs
\begin{equation}
\mathcal{F}<1.
\label{eq:Bconstraint1}
\end{equation}
For the solution~(\ref{eq:ansatz})--(\ref{Eqq2}), this bound reads
\begin{equation}
(3\eta+\beta \Lambda_\mathrm{bare})(\eta-\beta
\Lambda_\mathrm{bare})>0.
\label{eq:Bconstraint2}
\end{equation}
We have to keep in mind this constraint for the upcoming
stability analysis. The background solution~(\ref{Eqphi}) in the
physical frame $\tilde g_{\mu\nu}$ then recovers the same form as
the original background, namely:
\begin{align}
\label{eq:ansatzphys}
d\tilde s^2 &= -\tilde A(r)\, d\tilde t^2 +\frac{dr^2}{\tilde
B(r)} + r^2\left(d\theta^2 +\sin^2\theta\, d\phi^2\right),
\\
\tilde A(r) &= \tilde B(r)= 1- \frac{2G \tilde m}{r} - \frac{
\tilde \Lambda_\text{eff}}{3}\, r^2,
\\
\varphi &= \tilde q\left[\tilde t - \int\!\frac{\sqrt{1-\tilde
A(r)}}{ \tilde A(r)}\, dr\right],
\label{Eqphiphys}
\\
\tilde q &=\frac{q}{\sqrt{1-\mathcal{F}}}, \qquad \tilde m =
\frac{m}{1-\mathcal{F}}, \qquad \tilde \Lambda_\text{eff} =
\frac{\Lambda_\text{eff}}{1-\mathcal{F}}=\left(\frac{
\Lambda_\text{eff} +
\Lambda_\text{bare}}{3\Lambda_\text{eff}-\Lambda_\text{bare}}
\right) \Lambda_\text{eff},
\label{EqLambdaEffphys}
\end{align}
in the $(\tilde t, r)$ coordinate system with respect to the
rescaled parameters of the solution\footnote{Similarly to
Eq.~(\ref{Eqphi}), there actually exist two branches for the
scalar field, corresponding to a plus or minus sign in front of
the $r$ integral. In the physical frame, we keep only the minus
branch, so that this solution is mapped to a homogeneous and
expanding one in FLRW coordinates, see~\cite{Babichev:2016kdt}.
This minus sign actually also corresponds to a minus sign in the
Horndeski frame of Eq.~(\ref{Eqphi}).} $\tilde
\Lambda_\text{eff}$, $\tilde m$, $\tilde q$. Before studying the
stability, let us remark that the solution in the physical frame
is asymptotically de Sitter only for positive $\tilde
\Lambda_\mathrm{eff}$. Since this solution is meant to describe
our current Universe, we impose the positivity of $\tilde
\Lambda_\mathrm{eff}$. In terms of the Lagrangian parameters,
this translates as:
\begin{equation}
\eta
\beta(\eta-\beta\Lambda_\mathrm{bare})(3\eta+\beta\Lambda_\mathrm
{bare})<0,
\label{eq:exist2}
\end{equation}
to be combined with constraints~(\ref{eq:exist1})
and~(\ref{eq:Bconstraint2}). It is easy to check that the three
conditions together imply that the solution was also
asymptotically de Sitter in the original Horndeski frame, i.e.
that $\Lambda_\mathrm{eff}>0$. The above transformation can be
trivially extended to the stealth
solution~(\ref{eq:ansatz2})--(\ref{Eqphi2}). We hence recover the
announced result that the physical metrics are again black hole
solutions.

Let us now proceed with the perturbative analysis. Generically,
the theory~(\ref{eq:actionJ}) has one scalar degree of freedom,
and two polarizations of a massless spin-2 degree of freedom. We
first want to obtain the effective metric in which the scalar
mode propagates. To this end, we shall focus on a spherically
symmetric perturbation. If such a dynamical breather mode exists,
it necessarily corresponds to the scalar degree of freedom. We
perturb the metric and scalar field according to
\begin{align}
g_{\mu \nu}&=\bar{g}_{\mu \nu} +h_{\mu \nu},
\\
\varphi &= \bar{\varphi}+\chi,
\end{align}
where a bar denotes the background solution, and $h_{\mu \nu}$
and $\chi$ depend only on $t$ and $r$ since we look for a
spherically symmetric perturbation. Using the formalism developed
by Regge and Wheeler~\cite{Regge:1957td}, $h_{\mu \nu}$ can be
written in spherical coordinates as:
\begin{equation}
h_{\mu \nu}=\begin{pmatrix}
A(r) H_0(t,r) & H_1(t,r) & 0 & 0 \\
H_1(t,r) & H_2(t,r)/B(r) & 0 & 0 \\
0 & 0 & K(t,r) r^2 & 0\\
0 & 0 & 0 & K(t,r) r^2 \sin^2 \theta
\end{pmatrix}
,\end{equation}
where the $H_i$ and $K$ are free functions. Inserting these
perturbations into the action, we isolate the terms which are
quadratic in $h_{\mu \nu}$ and $\chi$. This gives the second
order perturbed action, that we can write as
\begin{equation}
\delta^{(2)}_\mathrm{s}S_\mathrm{J}=
\displaystyle\int{\mathrm{d}t\, \mathrm{d}r\, 4\pi
r^2\mathcal{L}^{(2)}_\mathrm{s}},
\end{equation}
where the factor $4\pi r^2$ corresponds to the trivial angular
integration, and $\mathcal{L}^{(2)}_\mathrm{s}$ is the Lagrangian
density from which we can extract the causal structure of the
perturbations. The subscript ``s'' stands for scalar, since we
choose to excite only a spherically symmetric mode. We can
simplify the calculations using the diffeomorphism invariance
generated by an infinitesimal vector $\xi^\mu$. In the new system
of coordinates $\hat{x}^\mu=x^\mu+\xi^\mu$, the metric and scalar
transform according to
\begin{align}
\hat{g}_{\mu \nu} &= g_{\mu \nu} -2\nabla_{(\mu}\xi_{\nu)},
\\
\hat{\varphi}&=\varphi-\partial_\mu\varphi \xi^\mu.
\end{align}
With a well-chosen $\xi^\mu$, we can in fact set $K$ and $\chi$
to zero. This completely fixes the gauge. Explicitly,
\begin{equation}
\xi^\mu=\left(\dfrac1q\left(\chi+\varphi'\dfrac{Kr}{2}\right),
-\dfrac{Kr}{2},0,0\right),
\end{equation}
a prime standing for a derivative with respect to $r$. In this
gauge, $\mathcal{L}^{(2)}_\mathrm{s}$ reads, after numerous
integrations by parts and using the background field equations,
\begin{equation}
\label{eq:H012Lag}
\mathcal{L}^{(2)}_\mathrm{s} = c_1 H_0
\dot{H}_2+c_2H_0'H_1+c_3H_0'H_2+c_4H_1\dot{H}_2+c_5H_0^2+c_6H_2^2
+c_7H_0H_2+c_8H_1H_2.
\end{equation}
Here a dot represents a time derivative, and all $c_i$ are
background coefficients with radial (but no time) dependence, the
detailed expression of which can be found in
appendix~\ref{sec:appmonop}. This three-field Lagrangian should
boil down to a Lagrangian depending on a single dynamical
variable. As a first step in this direction, it is easy to
eliminate $H_2$ since the associated field equation is algebraic
in $H_2$:
\begin{equation}
H_2 = -\dfrac{1}{2c_6}(-c_1\dot{H}_0-c_4 \dot{H}_1+c_3 H_0'+c_7
H_0+c_8H_1).
\end{equation}
Inserting back this expression in $\mathcal{L}^{(2)}_\mathrm{s}$,
we obtain,
\begin{equation}
\begin{split}
\mathcal{L}^{(2)}_\mathrm{s} &= \tilde{c}_1
\dot{H}_0^2+\tilde{c}_2H_0'^2+\tilde{c}_3H_0'\dot{H}_0+\tilde{c}
_4\dot{H}_1^2+\tilde{c}_5\dot{H}_0 H_1'+\tilde{c}_6 \dot{H}_0
\dot{H}_1+\tilde{c}_7H_0'H_1+\tilde{c}_8\dot{H}_0H_1
\\
&\quad+\tilde{c}_9H_0^2+\tilde{c}_{10}H_0H_1+\tilde{c}_{11}H_1^2,
\label{eq:H0H1Lag}
\end{split}
\end{equation}
where the $\tilde{c}_i$ coefficients are again given in
appendix~\ref{sec:appmonop} in terms of the $c_i$. A trickier
step is to trade $H_1$ and $H_0$ for a single variable, since the
associated field equations are differential equations, not
algebraic ones. To this end, we introduce an auxiliary field
$\pi_\mathrm{s}$ as a linear combination of $H_0$, $H_1$ and
their first derivatives:
\begin{equation}
\pi_\mathrm{s}=\dot{H}_0 + a_2 H_0' + a_3 \dot{H}_1 + a_4 H_1' +
a_5 H_0 + a_6 H_1,
\label{eq:pilinear}
\end{equation}
with some $a_i$ coefficients to be determined soon. The idea is
to introduce $\pi_\mathrm{s}$ at the level of the action, group
all the derivatives inside $\pi_\mathrm{s}$, and then to solve
for the algebraic equations giving $H_0$ and $H_1$ in terms of
$\pi_\mathrm{s}$. Therefore, we rewrite the Lagrangian as
\begin{equation}
\begin{split}
\mathcal{L}^{(2)}_\mathrm{s} &=
a_1[-\pi_\mathrm{s}^2+2\pi_\mathrm{s} (\dot{H}_0 + a_2 H_0' + a_3
\dot{H}_1 + a_4 H_1' + a_5 H_0 + a_6 H_1)]
\\
&\quad+ a_7 H_0^2+a_8 H_1^2+a_9 H_0 H_1.
\label{eq:piH0H1Lag}
\end{split}
\end{equation}
Variation of~(\ref{eq:piH0H1Lag}) with respect to
$\pi_\mathrm{s}$ ensures Eq.~(\ref{eq:pilinear}). Now, a simple
identification with Lagrangian~(\ref{eq:H0H1Lag}) allows us to
determine the $a_i$ in terms of the $\tilde{c}_i$. Again, these
coefficients are given in appendix~\ref{sec:appmonop}. Variation
of~(\ref{eq:piH0H1Lag}) with respect to $H_0$ and $H_1$ gives a
system of two linear equations, which we can easily solve to
write these two fields in terms of $\pi_\mathrm{s}$ and its
derivatives. We do not write down their expression here because
of their consequent length, but the procedure is
straightforward\footnote{The case of the stealth Schwarzschild
black hole~\cite{Babichev:2013cya} is more subtle. There, the two
equations determining $H_0$ and $H_1$ become linearly dependent
and the procedure cannot be applied. Section~\ref{sec:stealth} is
devoted to this particular case.}. At this point, we have
obtained a Lagrangian density in terms of a single variable
$\pi_\mathrm{s}$. We will examine its kinetic part only,
neglecting the potential associated to this degree of freedom and
thereby focusing on the causal structure. This kinetic part reads
\begin{equation}
\mathcal{L}_\mathrm{s;\:Kin}^{(2)}=
-\dfrac12(\mathcal{S}^{tt}\dot{\pi_\mathrm{s}}^2 +
2\mathcal{S}^{tr} \dot{\pi_\mathrm{s}} \pi_\mathrm{s}' +
\mathcal{S}^{rr} \pi_\mathrm{s}'^2),
\label{eq:piLag}
\end{equation}
with
\begin{align}
\mathcal{S}^{tt} &= \dfrac{c_1^2 c_3^2 c_4^2}{4c_2
\mathcal{D}}(-2c_4^2c_5+c_2c_1c_4'-c_2c_4c_1'-c_1c_4c_2'),
\\
\mathcal{S}^{rr} &= -\dfrac{c_1^2 c_3^2 c_4^2}{2
\mathcal{D}}(-c_3c_8+c_2c_6),
\\
\mathcal{S}^{tr} &= -\dfrac{c_1^2 c_3^2 c_4^2}{4
\mathcal{D}}(-c_4c_7+c_1c_8),
\\
\begin{split}
\label{Ddef}
\mathcal{D} &= c_6^2 \Bigl\{2(-c_3 c_8+c_2c_6)
(-c_2c_4c_1'+c_1c_2c_4'-c_1c_4c_2')
\\
&\quad+\bigl[4c_3c_8c_5+c_2(c_7^2-4c_6c_5)\bigr]c_4^2
-2c_2c_4c_7c_1c_8+c_2c_1^2c_8^2\Bigr\}.
\end{split}
\end{align}
Alternatively, we can remark that the scalar mode propagates to
linear order in the given black hole background~(\ref{eq:ansatz})
with an effective two-dimensional metric $\mathcal{S}^{-1}_{\mu
\nu}$:
\begin{equation}
\mathcal{L}_\mathrm{s;\:Kin}^{(2)}= - \dfrac12 \mathcal{S}^{\mu
\nu} \partial_\mu \pi_\mathrm{s} \partial_\nu \pi_\mathrm{s}.
\end{equation}
We can read from Eq.~(\ref{eq:piLag}) the inverse metric:
\begin{equation}
\mathcal{S}^{\mu \nu}=\begin{pmatrix}
\mathcal{S}^{tt} & \mathcal{S}^{tr} \\
\mathcal{S}^{tr} & \mathcal{S}^{rr}
\end{pmatrix}
,\end{equation}
and the metric itself:
\begin{equation}
\mathcal{S}^{-1}_{\mu \nu}=\dfrac{1}{\mathcal{S}^{tt}
\mathcal{S}^{rr}-(\mathcal{S}^{tr})^2}\begin{pmatrix}
\mathcal{S}^{rr} & -\mathcal{S}^{tr} \\
-\mathcal{S}^{tr} & \mathcal{S}^{tt}
\end{pmatrix}.
\end{equation}
{}From this last object, we can determine the hyperbolicity
condition, the propagation speeds, and all the information we
need for the causal structure of the scalar mode. The
hyperbolicity condition for instance reads
\begin{equation}
(\mathcal{S}^{tr})^2-\mathcal{S}^{tt}\mathcal{S}^{rr}>0.
\label{eq:hypscal}
\end{equation}
The speed of a wave moving towards or away from the origin is
then given by
\begin{equation}
c_\mathrm{s}^\pm=\dfrac{\mathcal{S}^{tr} \pm
\sqrt{(\mathcal{S}^{tr})^2-\mathcal{S}^{tt}\mathcal{S}^{rr}}}
{\mathcal{S}^{tt}}.
\end{equation}
The hyperbolicity condition ensures that these propagation speeds
are well defined. At any given point, $c_\mathrm{s}^+$ and
$c_\mathrm{s}^-$ generate the scalar causal cone. Finally, one
needs to know where the interior of the cone is located. This can
be easily determined by checking whether a given direction (for
instance the one generated by the vector $\partial_t$) is time or
space-like with respect to the metric $\mathcal{S}^{-1}_{\mu
\nu}$.

A similar analysis must be carried out for the spin-2 mode. It
was actually already realized by Ogawa \textit{et al.}
in~\cite{Ogawa:2015pea}. They studied odd-parity perturbations,
which cannot correspond to a scalar degree of freedom ---~the
latter always has even parity. Hence odd-parity perturbations
correspond to one of the two spin-2 polarizations. We checked the
calculations of Ogawa \textit{et al.}, and we are in full
agreement as to the quadratic Lagrangian derived in their paper.
For brevity, we only reproduce the final result here, applied to
the solution~(\ref{eq:ansatz})--(\ref{Eqq2}); the gravity
perturbations propagate in a two-dimensional effective metric
$\mathcal{G}^{-1}_{\mu \nu}$, which essentially coincides with
the physical metric $\tilde g_{\mu\nu}$, as given by
Eq.~(\ref{eq:disformal}):
\begin{equation}
\mathcal{G}^{-1}_{\mu\nu}= \dfrac{\Lambda_\mathrm{eff}}
{\Lambda_\mathrm{bare}+\Lambda_\mathrm{eff}}\, \tilde g_{\mu\nu}.
\label{eq:Gmetric}
\end{equation}
The two metrics are related by a constant conformal factor.
Therefore, they have identical causal structure at \textit{any}
point of spacetime, provided that the conformal factor is
positive. When this is the case, matter and gravitons propagate
exactly the same way; it is enough to analyze the overlap
conditions of two of the three causal cones, say the matter and
scalar one. On the contrary, if the above conformal factor is
negative, the cone of the graviton is exactly complementary to
the matter one and they have no overlap nor common exterior. We
are therefore led to impose that
\begin{equation}
\Lambda_\mathrm{eff}(\Lambda_\mathrm{bare}+\Lambda_\mathrm{eff})>0,
\label{eq:samemetric1}
\end{equation}
i.e. in terms of the Lagrangian parameters:
\begin{equation}
\eta(\eta-\beta\Lambda_\mathrm{bare})>0,
\label{eq:samemetric2}
\end{equation}
The hyperbolicity condition coming from
$\mathcal{G}^{-1}_{\mu\nu}$, or equivalently $\tilde g_{\mu\nu}$,
reads:
\begin{equation}
(\mathcal{G}^{tr})^2-\mathcal{G}^{tt}\mathcal{G}^{rr}>0,
\label{eq:hypgrav}
\end{equation}
and the speeds of inwards/outwards moving gravitons are given by
\begin{equation}
c_\mathrm{g}^\pm=\dfrac{\mathcal{G}^{tr} \pm \sqrt{(\mathcal{G}^{tr})^2
-\mathcal{G}^{tt}\mathcal{G}^{rr}}}{\mathcal{G}^{tt}}.
\end{equation}
In a nutshell, we have found the effective metrics in which
gravitons and massless scalar propagate and they are given by
$\mathcal{G}^{\mu \nu}$ and $\mathcal{S}^{\mu \nu}$ respectively.
The Schwarzschild-de Sitter solution is well defined in both
Einstein and physical frames provided Eqs.~(\ref{eq:exist1}),
(\ref{eq:Bconstraint2}) and (\ref{eq:exist2}). Additionally,
under condition~(\ref{eq:samemetric2}), gravitons propagate in
the exact same metric as matter. In particular, the speed of
gravitational waves is identical to the speed of light in a
strongly curved background.

\subsection{Homogeneous solutions: a stability window}

We will first apply the above analysis to de Sitter solutions,
that is solution~(\ref{eq:ansatz})-(\ref{Eqq2}) with $m=0$. Of
course, in this case, the analysis presented above is not
strictly necessary, but it allows us to cross check our results
with cosmological perturbation theory. In particular, we arrive
at the same conclusion as~\cite{Appleby:2011aa} for the
model~(\ref{eq:actionJ}) with vanishing $\Lambda_\text{bare}$:
There is no stable homogeneous configuration. However, switching
on a non trivial $\Lambda_\text{bare}$, the hyperbolicity
conditions~(\ref{eq:hypscal}) and~(\ref{eq:hypgrav}) read
respectively:
\begin{align}
(3 \beta\Lambda_\text{bare}+\eta)(\eta-\beta\Lambda_\text{bare})&<0,
\label{eq:hypscal2}
\\
(3\eta+\beta\Lambda_\text{bare})(\eta-\beta\Lambda_\text{bare})&>0.
\label{eq:hypgrav2}
\end{align}
These two conditions must be supplemented with the fact that the
graviton and scalar cones have a non-empty intersection and a
common exterior. Again, compatibility with the matter causal cone
will follow automatically, since $\tilde g^{\mu\nu}$ and
$\mathcal{G}^{\mu\nu}$ are conformally related, with a positive
factor provided Eq.~(\ref{eq:samemetric2}). It is enough to check
the orientation of the cones at $r=0$, since the solution under
analysis is homogeneous. If the cones have compatible
orientations at $r=0$, this will remain true everywhere else. The
calculation is then particularly simple, since $\mathcal{S}^{tr}$
and $\mathcal{G}^{tr}$ vanish at $r=0$, meaning that the cones
are either aligned (and symmetric around the $t$ axis) or
inclined at ninety degrees. As soon as the hyperbolicity
condition~(\ref{eq:hypscal2}) for the scalar and the
constraint~(\ref{eq:samemetric2}) are satisfied, the $t$ axis is
contained in the cone associated to $\mathcal{S}^{-1}_{\mu \nu}$.
We therefore need the graviton cone to contain also the $t$ axis.
This is the case if
\begin{equation}
\eta (3 \eta+\beta \Lambda_\text{bare})>0.
\label{eq:goodorient}
\end{equation}
Thus, there are in total seven conditions to fulfill for
stability and existence of the solution: Eqs.~(\ref{eq:exist1}),
(\ref{eq:Bconstraint2}), (\ref{eq:exist2}),
(\ref{eq:samemetric2}), (\ref{eq:hypscal2}), (\ref{eq:hypgrav2})
and (\ref{eq:goodorient}). They actually define an non-empty
subspace of the parameter space. The cosmological solution is
stable if and only if
\begin{align}
\mathrm{either}~\eta&>0,~\beta
<0~\mathrm{and}~\dfrac{\Lambda_\text{bare}}{3}
<-\dfrac{\eta}{\beta}<\Lambda_\text{bare},
\label{eq:range1}
\\
\mathrm{or}~\eta&<0,~\beta>0~\mathrm{and}~\Lambda_\text{bare}
<-\dfrac{\eta}{\beta}<3\Lambda_\text{bare}.
\label{eq:range2}
\end{align}
In the following section, we give an example of parameters that
fulfill this criterion. Let us stress that the above restrictions
prevent one from using the theory~(\ref{eq:actionJ}) as a
self-tuning model. Indeed, the above equations tell us that the
effective cosmological constant has to be of same magnitude as
the bare one. Rewriting these conditions in terms of the observed
$\tilde\Lambda_\text{eff}$, we obtain
\begin{align}
\mathrm{either}~\eta&>0,~\beta<0~\mathrm{and}~\Lambda_\text{bare}
< \tilde\Lambda_\text{eff},
\\
\mathrm{or}~\eta&<0,~\beta>0~\mathrm{and}~\Lambda_\text{bare} <
\tilde\Lambda_\text{eff} <\frac{3}{2}\,\Lambda_\text{bare}.
\end{align}
Again, this means that self-tuning is impossible in this specific
model, since the observed cosmological constant must always be
larger than the bare one.

\subsection{Black holes in de Sitter: example of a stable
configuration}

Our analysis is fully relevant when the solution no longer
describes a homogeneous cosmology, but rather a black hole
embedded in such a cosmology. The three conditions for the
background solution to exist and the frame transformation to be
well-defined, Eqs.~(\ref{eq:exist1}), (\ref{eq:Bconstraint2}),
(\ref{eq:exist2}), do not depend on the presence of a mass
$m\neq0$. Therefore, they remain identical when a black hole is
present. Additionally, the condition~(\ref{eq:samemetric2}) for
$\tilde g_{\mu\nu}$ and $\mathcal{G}^{-1}_{\mu\nu}$ to have
compatible orientations is unchanged. Provided
Eq.~(\ref{eq:samemetric2}), photons coupled to $\tilde
g_{\mu\nu}$ and gravitons will travel with the exact same speed,
even in a highly curved background. This is a very positive
feature of the theory~(\ref{eq:actionJ}). Indeed, comparing the
speed of gravitational and electromagnetic waves with arbitrary
accuracy cannot rule out this model.

\begin{figure}
\centering
\includegraphics[width=\textwidth]{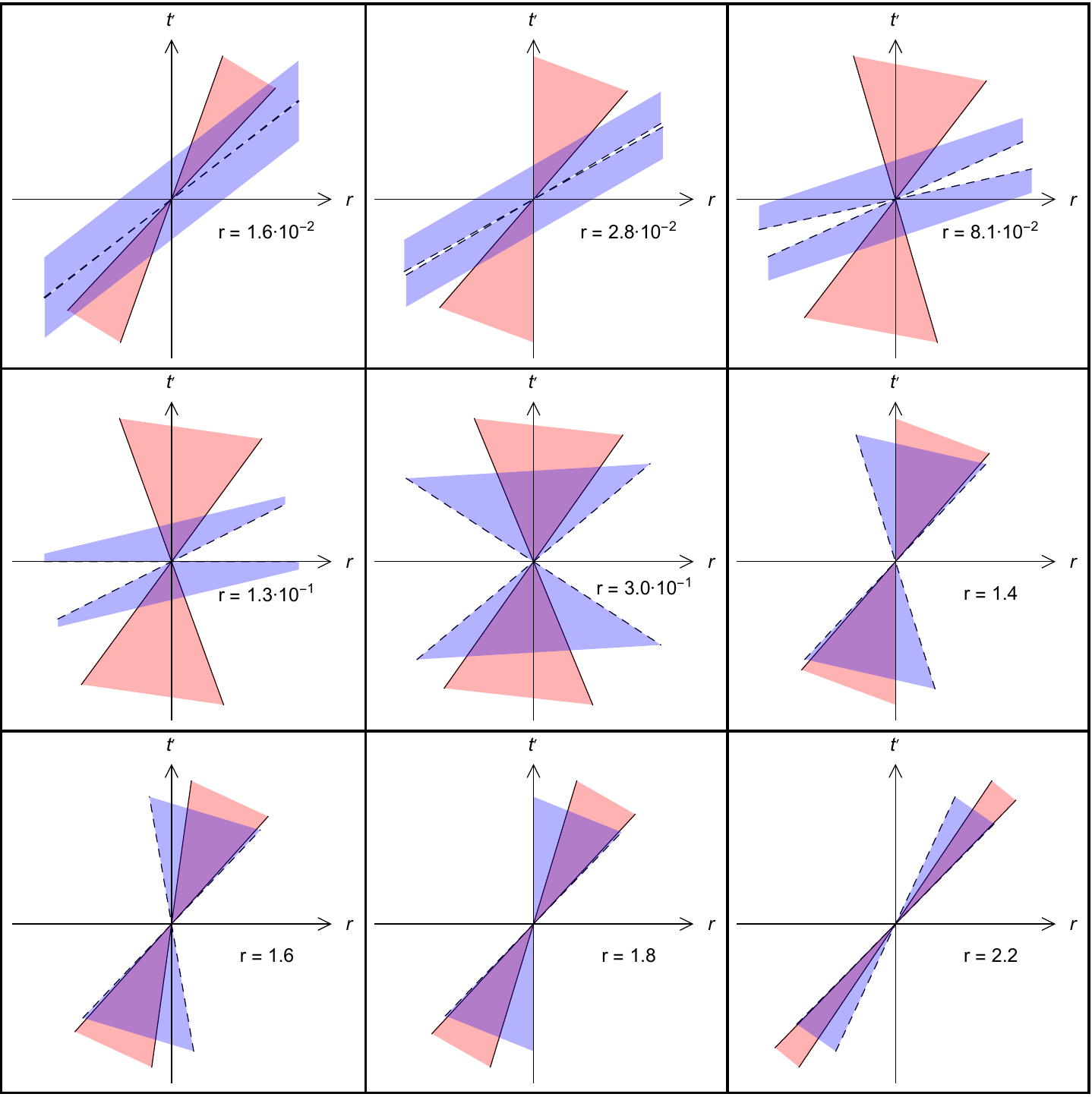}
\caption{The scalar and graviton/matter causal cones in
Schwarzschild-de Sitter geometry, respectively in dashed blue and
plain red. The parameters of the Lagrangian are chosen so that
the associated cosmological solution is stable:
$\eta=\frac{1}{2}$, $\beta=-1$, $\zeta=1$,
$\Lambda_\text{bare}=1$ in Planck units. The radius $r$ varies
between the black hole horizon located at $r\simeq 9.4\cdot
10^{-3}$ and the cosmological horizon at $r\simeq2.4$. For this
set of parameters, the graviton cone always lies inside the
scalar cone; as a consequence, they have compatible orientations.
In this plot, the time coordinate $t'$ has been rescaled with
respect to the original one, so that the causal cone associated
to the unphysical metric $g_{\mu\nu}$ corresponds to lines at
$\pm45^\circ$.}
\label{fig:bhsoundcone}
\end{figure}

On the other hand, the expressions of $\mathcal{G}^{-1}_{\mu
\nu}$ and $\mathcal{S}^{-1}_{\mu \nu}$ become very complicated
with a non-vanishing black hole mass. It is still possible to
prove that the hyperbolicity conditions for both
$\mathcal{S}^{-1}_{\mu \nu}$ and $\mathcal{G}^{-1}_{\mu \nu}$ are
not modified with respect to the de Sitter case. They are again
given by Eqs.~(\ref{eq:hypscal2}), (\ref{eq:hypgrav2}). To ensure
the compatibility of orientation between the scalar cone and the
graviton one is however more tricky. We checked numerically that
the condition~(\ref{eq:goodorient}) for these two cones to be
compatible in the de Sitter case leads to compatible cones also
when the mass parameter $m$ is switched on. That is, for
parameters in the range~(\ref{eq:range1}) or~(\ref{eq:range2}),
the scalar cone seems to have a compatible orientation with the
graviton cone even close to the black hole horizon. This remains
true for arbitrary mass of the black hole (as long as the black
hole horizon remains smaller than the cosmological horizon).
Figure~\ref{fig:bhsoundcone} provides an illustrative example of
this numerical check, for a given set of parameters that falls in
the range allowed by the corresponding de Sitter solution. In
this case, the cones have compatible orientations everywhere.
Remarkably, the scalar causal cone entirely opens up when
approaching the black hole horizon, without becoming pathological.

Let us stress here why Ref.~\cite{Ogawa:2015pea} would have
claimed that the situation exhibited in
Fig.~\ref{fig:bhsoundcone} is unstable, in the light of the
discussion of Sec.~\ref{Sec2}. In this paper, the graviton metric
components $\mathcal{G}^{tt}$ and $\mathcal{G}^{rr}$ were
required to be negative and positive respectively. It was proven,
however, that the product $\mathcal{G}^{tt}\mathcal{G}^{rr}$ is
always positive in the vicinity of a horizon.
Figure~\ref{fig:bhsoundcone} shows that, indeed, the $t$ axis
``leaves'' the causal cone of the graviton (red cone), close to
the event \textit{and} cosmological horizon, while the $r$ axis
remains in the exterior of the cone. This makes the quantity
$\mathcal{G}^{tt}\mathcal{G}^{rr}$ positive close to the horizon,
and the associated Hamiltonian unbounded by below. However, our
analysis so far clearly shows that it does not signal an
instability in any way.

{}From Fig.~\ref{fig:bhsoundcone}, one can additionally draw
conclusions on the case where matter couples to $g_{\mu\nu}$,
rather than $\tilde g_{\mu\nu}$. Of course, matter was chosen to
couple to $\tilde g_{\mu \nu}$ on physical grounds. However, in a
generic scalar-tensor theory where no relation is assumed between
the propagation of gravity and light, one has to investigate
these three different cones. The time coordinate $t'$ in
Fig.~\ref{fig:bhsoundcone} is rescaled with respect to $t$ in
such a way that the causal cone associated with $g_{\mu\nu}$ is
at $\pm45^\circ$, with a timelike $t'$ direction (that is, the
$t'$ axis lies inside the cone of $g_{\mu\nu}$). Even in this
more restrictive situation, the plots of
Fig.~\ref{fig:bhsoundcone} show that the three cones would
actually be compatible, and the solution would be stable.

\subsection{A special case: stealth Schwarzschild black hole}
\label{sec:stealth}

As mentioned above, there exists an exact asymptotically flat
Schwarzschild solution when $\eta$ and $\Lambda_\text{bare}$
vanish, with a non-trivial scalar
profile~(\ref{eq:ansatz2})--(\ref{Eqphi2}). In this case, the
parameter $q$ is no longer related to the coupling constants of
the action and is in fact a free parameter. We should also
emphasize that solution~(\ref{eq:ansatz2})--(\ref{Eqphi2}) is the
unique static and spherically symmetric solution with a linearly
time dependent scalar field and $\eta=\Lambda_\text{bare}=0$. The
procedure for determining the effective metric of scalar
perturbations, described in Sec.~(\ref{sec:effmetrics}), breaks
down for this background. It is not possible to carry on after
Eq.~(\ref{eq:H0H1Lag}), and to express the fields $H_0$ and $H_1$
in terms of $\pi_\mathrm{s}$. The reason is that for the stealth
Schwarzschild solution $H_0$ and $H_1$ cannot be simultaneously
expressed in terms of the master variable $\pi_s$ from the
Lagrangian introduced in~(\ref{eq:piH0H1Lag}). Therefore, we need
to find another way of extracting the scalar mode from the
second-order Lagrangian~(\ref{eq:H012Lag}) which now reads
\begin{equation}
\label{eq:H012Lagbis}
\mathcal{L}^{(2)}_\mathrm{s} = c_1 H_0
\dot{H}_2+c_2H_0'H_1+c_3H_0'H_2+c_4H_1\dot{H}_2
+c_6H_2^2+c_7H_0H_2+c_8H_1H_2,
\end{equation}
as $c_5=0$ for the relevant background. The equation of motion
for $H_2$ following from~(\ref{eq:H012Lag}) is algebraic in terms
of $H_2$, so as before we can find $H_2$ in terms of $H_0$ and
$H_1$:
\begin{equation}
\label{H2}
H_2 = -\dfrac{1}{2c_6}(-c_1\dot{H}_0-c_4 \dot{H}_1+c_3 H_0'+c_7
H_0+c_8H_1).
\end{equation}
Substituting~(\ref{H2}) in~(\ref{eq:H012Lagbis}) and rearranging
terms, we can write~(\ref{eq:H012Lagbis}) as
\begin{equation}
\label{eq:H01Lag}
\mathcal{L}^{(2)}_\mathrm{s} = a_1(\dot{H}_0 + a_2 H_0' + a_3
\dot{H}_1 + a_5 H_0 + a_6 H_1)^2
+ a_7 H_0^2+a_8 H_1^2+a_9 H_0 H_1,
\end{equation}
where the coefficients $a_i$ are given in the
Appendix~\ref{sec:appmonop}. We now introduce new variables
$x(t,r)$ and $y(t,r)$ grouping together the time and space
derivatives:
\begin{equation}
\begin{split}
H_1 & = \frac{c_1}{c_4}(x-\frac{c_1}{c_3}y),
\\
H_0 &= \frac{c_1}{c_3}y.
\end{split}
\end{equation}
Indeed, the Lagrangian~(\ref{eq:H01Lag}) then takes the form
\begin{equation}
\label{Lmaster1}
\mathcal{L}^{(2)}_\mathrm{s} = \mathcal{P}^2 + \mathcal{A}y^2 +
\mathcal{B}xy + \mathcal{C}x^2,
\end{equation}
where
\begin{equation}
\mathcal{P}= \dot{x}-y'+\tilde{a}_1 x + \tilde{a}_2 y,
\end{equation}
and
\begin{equation}
\begin{split}
\tilde{a}_1 &= \frac{2 c_2 c_6-c_3 c_8}{c_3 c_4},\\
\tilde{a}_2 &= \frac{c_4 c_1 c_3'-c_3 c_4 c_1'+c_8 c_1^2-c_4 c_7
c_1}{c_1 c_3 c_4},\\
\mathcal{A} &= \frac{c_1^2 \left[c_4 \left(c_2 c_1'+c_1 c_2'+2
c_4 c_5\right)-c_1 c_2 c_4'\right]}{2 c_3^2 c_4^2},\\
\mathcal{B} &= \frac{c_1^2 c_2 \left(c_1 c_8-c_4
c_7\right)}{c_3^2 c_4^2},\\
\mathcal{C} &= \frac{c_1^2 c_2 \left(c_2 c_6-c_3
c_8\right)}{c_3^2 c_4^2}.\\
\end{split}
\end{equation}
Variation of~(\ref{Lmaster1}) with respect to $y$ yields the constraint
\begin{equation}
\label{constraintxy}
2\mathcal{P}' + 2\mathcal{A}y+\mathcal{B}x = 0.
\end{equation}
The above constraint~(\ref{constraintxy}) contains $y''$, $y'$,
$\dot{x}'$ and it may be seen as an equation which determines $y$
in terms of $x$ and its derivatives. To find $y$
from~(\ref{constraintxy}), the use of nonlocal (in space)
operators is in general required. For our purposes, however, we
do not need to know the exact expression of $y$ in terms of $x$,
since we are only interested in the absence of ghost and gradient
instabilities. This means that we focus on higher derivative
terms, i.e. we neglect $\sim x$ with respect to $\sim \dot{x}$ or
$\sim x'$ as well as $\sim y$ with respect to $y'$. With this
approximation in mind, equation~(\ref{constraintxy}) becomes
\begin{equation}
\dot{x}' - y''=0,
\end{equation}
which after integration over $r$ and setting to zero the
integration constant yields
\begin{equation}
\label{constraintxy2}
\dot{x}=y'.
\end{equation}
By the same token, Eq.~(\ref{constraintxy}) shows that the term
$\mathcal{P}^2$ in~(\ref{Lmaster1}) is of lower order in
derivatives in comparison with the last three terms, because
from~(\ref{constraintxy}) one can see that $\mathcal{P}$ is of
lower order compared to $x$ and $y$. As a consequence, to the
leading order in derivatives, the Lagrangian~(\ref{Lmaster1}) is
\begin{equation}
\label{Lmaster2}
\mathcal{L}_\mathrm{s;\:Kin}^{(2)} = \mathcal{A}y^2 +
\mathcal{B}xy + \mathcal{C}x^2,
\end{equation}
where the subscript ``Kin'' stresses that only higher order terms
(kinetic part) are left in the Lagrangian. We then introduce
$\tilde{\pi}$ as
\begin{equation}
\label{xpi}
x = \tilde\pi',
\end{equation}
and from (\ref{constraintxy2}), we easily obtain
\begin{equation}
\label{ypi}
y = \dot{\tilde\pi},
\end{equation}
where we set to zero the integration constant. Finally,
substituting~(\ref{xpi}) and~(\ref{ypi}) in~(\ref{Lmaster2}), we
find the kinetic part of the Lagrangian for the scalar
perturbations:
\begin{equation}
\label{Lmaster3}
\mathcal{L}_\mathrm{s;\:Kin}^{(2)} =-\frac12
\left(\tilde{\mathcal{S}}^{tt}\dot{\tilde\pi}^2 +2
\tilde{\mathcal{S}}^{tr}\dot{\tilde\pi}\tilde\pi' +
\tilde{\mathcal{S}}^{rr}\tilde\pi'^2\right),
\end{equation}
where
\begin{equation}
\label{SABC}
\tilde{\mathcal{S}}^{tt} = - 2\mathcal{A}, \quad
\tilde{\mathcal{S}}^{tr} = - \mathcal{B}, \quad
\tilde{\mathcal{S}}^{rr} = - 2\mathcal{C}.
\end{equation}
One can obtain the same result for the de Sitter black hole by
following the above method rather than~(\ref{eq:piLag}).
Indeed, first of all, the hyperbolicity condition
for~(\ref{Lmaster3}) reads
\begin{equation}
\label{hyperbolicity}
D \equiv \tilde{\mathcal{S}}^{tt}\tilde{\mathcal{S}}^{rr}
-(\tilde{\mathcal{S}}^{tr})^2<0.
\end{equation}
The explicit expression for $D$ in terms of $c_i$ is given by
\begin{equation}
\label{Determ}
D = - \frac{c_1^4 c_2 \left\{c_2 \left(c_4 c_7-c_1
c_8\right){}^2-2 \left(c_2 c_6-c_3 c_8\right) \left[c_4
\left(c_2 c_1'+c_1 c_2'+2 c_4 c_5\right)-c_1 c_2
c_4'\right]\right\}}{c_3^4 c_4^4}.
\end{equation}
One can also verify that
\begin{equation}
D = -\frac{c_1^4 c_2}{c_3^4 c_4^4 c_6^2} \mathcal{D},
\end{equation}
where $\mathcal{D}$ is defined in (\ref{Ddef}). In terms of $D$,
the hyperbolicity condition found in~(\ref{eq:hypscal}) reads,
\begin{equation}
\frac{c_1^8}{16 c_6^4 D}<0.
\end{equation}
As long as $D<0$, i.e. the hyperbolicity
condition~(\ref{hyperbolicity}) is satisfied for $\tilde{\pi}$,
the hyperbolicity condition is also satisfied for $\pi$.
Moreover, for $D<0$ the variables $\tilde{\pi}$ and $\pi$ and the
kinetic matrices for $\tilde{\pi}$ and $\pi$ are related as,
\begin{equation}
\label{Srelation}
\mathcal{S}^{ab} = -\frac{c_1^4}{4 c_6^2 D}\,
\tilde{\mathcal{S}}^{ab},\quad
\pi = \frac{2 c_6}{c_1^2} \sqrt{-D}\,\tilde{\pi}.
\end{equation}
where indices $a$ and $b$ are either $t$ or $r$.

The advantage of the Lagrangian~(\ref{Lmaster3}) obtained here is
that it also allows to treat the case of stealth Schwarzschild
black hole, for which the method of Sec.~\ref{sec:effmetrics}
fails. Indeed, for the stealth solution it turns out that
$\mathcal{D}=D=0$ (in other words, $H_0$ and $H_1$ are linearly
dependent). However, the kinetic matrix
$\tilde{\mathcal{S}}^{ab}$ remains finite, see~(\ref{SABC}),
while the kinetic matrix $\mathcal{S}^{ab}$ diverges, as it can
be seen from~(\ref{Srelation}).

For the Lagrangian~(\ref{Lmaster3}), the vanishing determinant of
the kinetic matrix means that the equation of motion is parabolic
(for all $r$). {\it Per se}, this fact does not necessarily mean
that the perturbations are pathological on the considered
background. For instance, in the case of the k-essence Lagrangian
$\mathcal{L}= G_2(X)$, for solutions where $dG_{2}/dX=0$ with
{\it timelike} $\nabla^\mu\varphi$, the perturbations behave as
dust, i.e., they are governed by a wave equation with
$c_\text{s}^2=0$. The determinant of the kinetic matrix in this
case is also zero, since only the $tt$ component of the kinetic
matrix is non vanishing. For the stealth solution
(\ref{eq:ansatz2})--(\ref{Eqphi2}), the kinetic matrix reads
\begin{equation}
\label{kinmatrs}
\tilde{\mathcal{S}}^{ab} \propto
\begin{pmatrix}
\dfrac{\mu r}{(\mu-r)^2} & \dfrac{\sqrt{\mu r}}{\mu-r} \\
\dfrac{\sqrt{\mu r}}{\mu-r} & 1
\end{pmatrix},
\end{equation}
where we defined $\mu=2Gm$. Notice that, for $r\gg \mu$, all the
terms of~(\ref{kinmatrs}) apart from $\tilde{\mathcal{S}}^{rr}$
tend to zero. The global factor of Eq.~(\ref{kinmatrs}) may have
any sign, depending on the parameters of the model and the
(arbitrary) value of $q$ in (\ref{Eqphi2}). When this global
factor is negative, the dynamics of the perturbations indeed
corresponds to dust (i.e., a vanishing velocity, similarly to the
example of k-essence described above), but the infinitely thin
cone of propagation tends towards the $r$ axis. This, together
with the fact that the graviton cone has a ``usual'' behavior at
$r\to \infty$, makes the stealth solution pathological. It
corresponds to a limit of panels (i) and (j) of
Fig.~\ref{fig:CausalCones} when the dashed (blue) cone is
infinitely thin. On the other hand, for a positive global factor
in Eq.~(\ref{kinmatrs}), the scalar dynamics corresponds to the
limit of panels (c) and (d) of Fig.~\ref{fig:CausalCones} when
the dashed (blue) cone totally opens, i.e., its sound velocity is
infinite. In that case, the scalar field is no longer a
propagating degree of freedom.

\section{Conclusions}
\label{sec:conc}

In this paper, we have studied stability criteria for solutions
in modified gravity theories. We then applied these criteria to
establish the stability of certain hairy black holes whose hair
is supplied by a dark energy scalar field
\cite{Babichev:2013cya,Charmousis:2015aya,Babichev:2016kdt,%
Babichev:2016rlq}.

Throughout this study, we focused on scalar-tensor theories, but
the tools we have developed, as well as the stability criteria
concerning Hamiltonian densities, are generically applicable in
modified gravity theories. The starting ingredient for the
applicability of our tools are multiple gravitational modes, a
clear characteristic of theories going beyond GR. In order to
treat the problem consistently, we have formulated the notion of
causal cones, each of which is associated to a healthy
propagating degree of freedom. Indeed, the local existence of
well-defined causal cones permits us to determine the healthy
propagation of modes about an effective background solution. We
saw that, unlike standard lore, the Hamiltonian densities
associated to each of the modes do not suffice to exhibit an
instability. The failure of the Hamiltonian criterion, in more
complex background metrics, is due to the fact that it is not a
scalar quantity. Each Hamiltonian density, associated to a
propagating mode, depends on the particular coordinate system one
is using. So although a Hamiltonian density which is bounded from
below signals that the mode is stable, the converse is not true.
Namely, a Hamiltonian density found to be unbounded from below in
some coordinate system is inconclusive on instability. One may
find a coordinate transformation rendering the Hamiltonian
bounded from below as we saw explicitly in Sec.~2.

Standard lore is recovered only for a class of ``good''
coordinate systems that are defined with respect to all the
causal cones present in the system: scalar, graviton, matter,
etc. Namely, these ``good'' coordinate systems exhibit a timelike
coordinate common to all causal cones and spacelike coordinates
for all causal cone exteriors. If such a coordinate system
exists, then the Hamiltonian is indeed bounded from below and the
modes are well behaved, propagating in a timelike direction with
a hyperbolic operator. If not, then indeed the Hamiltonian for at
least one of the modes is always unbounded from below, and the
said mode presents a gradient or ghost instability.

The subtlety arises due to the complexity of the background
solution. Indeed, the key point for our examples here is that the
background scalar is space and time dependent. Then the causal
cones can tilt and open up as we approach the horizon (event or
cosmological). As a result, the original time (space) coordinate
of the background metric may ``leave'' the interior (exterior) of
a causal cone associated to some mode. This can lead to a
misinterpretation of the Hamiltonian density associated to the
initial coordinates, which ``leave'' the causal cone of the mode
in question.
We should emphasize that the failing Hamiltonian stability
criterion is not due to the mixing of modes, as illustrated by
the simple k-essence example of Sec.~\ref{Sec2}. We may also
consider the case of a theory including a $G_3$ Horndeski term
(the DGP term) where the mixing and demixing of modes has been
completely resolved \cite{Babichev:2012re} for an arbitrary
background\footnote{This result is not known for $G_4$ theories
for example.}. In such a model were found self-accelerating
vacua for the so-called Kinetic Gravity Braiding (KGB) model
\cite{Deffayet:2010qz} (see also \cite{Babichev:2016fbg}).
In a standard FLRW coordinate system, where the dark energy
scalar depends purely on cosmological time, such KGB
self-accelerating solutions generically give (depending on the
coupling constants of the theory) a stable vacuum, with an
associated Hamiltonian which is bounded from below. When one
considers the precise same stable vacua in a spherical coordinate
system, where the metric is static but the scalar field now
depends both on space and time, the same Hamiltonian density can
be found to be unbounded from below.
This, as we emphasized, is an artifact of a bad use of a
coordinate system (here spherical) whereas the FLRW coordinates
are indeed ``good'' (satisfying the causal cone criteria, its
cosmological time remaining notably within the causal cones).
This example demonstrates that misinterpretations related to
Hamiltonians are not due to mixing of modes but, crucially, to
the background depending (or not) on multiple coordinates. It is
for this reason that we do not encounter problems with the
Hamiltonian in FLRW systems for example or with static black
holes (with a static scalar field). One therefore expects our
analysis to be relevant for backgrounds (in modified gravity)
with lesser symmetry, for example stationary backgrounds
involving rotating black holes. For stationary backgrounds for
example, the $\theta$ and $r$ dependent background effective
metrics may again tilt and open up as we approach horizons. Also
clearly our analysis could be used in vector-tensor theories
where similar black holes to the one studied here have been found
\cite{Chagoya:2016aar,Minamitsuji:2016ydr,Babichev:2017rti}.
In any case, let us emphasize that there always exist ``bad''
coordinates in which the Hamiltonian density of a stable solution
appears unbounded by below. It suffices that its time direction
be outside at least one of its causal cones, or that one of its
spatial directions be inside one of them. For instance, a mere
exchange of $t$ and $x$ creates such a spurious pathology,
whereas the physics is obviously unchanged. As usual in GR, one
should never trust coordinate-dependent quantities.

We also underlined that when there exists a ``good'' coordinate
system in which the total Hamiltonian density is bounded by
below, then it may also be computed in other coordinate systems,
but it no longer corresponds to the mere Hamiltonian. It becomes
a linear combination of the energy and momentum densities, whose
spatial integrals over the whole system are all conserved. In
other words, the stability of the solution is still guaranteed by
the boundedness by below of a conserved quantity, but this is no
longer the mere energy which plays this role. The conditions for
the existence of such a ``good'' coordinate system, i.e., for
stability, are written in a covariant form in
Eqs.~(\ref{eq:gUU})--(\ref{eq:positiveT00}).

Using the above tools, we have corrected a misinterpretation
\cite{Ogawa:2015pea,Takahashi:2015pad,Takahashi:2016dnv,%
Kase:2018voo} in the literature about the said instability of a class
of hairy black holes. It is true, as stated
in~\cite{Ogawa:2015pea}, that the Hamiltonian for the graviton is
always unbounded by below in Schwarzschild's coordinates when
approaching a horizon. However, at the same time, the graviton
causal cone remains compatible (and may even coincide) with the
matter causal cone under some conditions on the parameters
defining the model. In other words, there exists coordinate
systems where the Hamiltonian for the graviton is bounded by
below. We completed this stability analysis by computing the
scalar causal cone, thanks to the study of $\ell=0$
perturbations. Again, we found that there exists a domain of
parameters where the three causal cones share a common time and a
common spacelike hypersurface. Hence, the class of hairy black
holes studied here, that is quite generically encountered in
Horndeski and beyond Horndeski theories
\cite{Babichev:2016rlq,Babichev:2016kdt}, is free of ghost and
gradient instability pathologies for a given range of parameters
of the model. This is an important result considering the absence
of stable hairy black holes in gravitational physics (see for
example
\cite{Bekenstein:1975ts, Bronnikov:1978mx, Harper:2003wt,%
Volkov:1998cc} for two celebrated cases). We have demonstrated
this result for a particular Horndeski theory but the result will
be similar for other cases and we even expect some cases will
allow for self-tuning properties \cite{Babichev:2017lmw}. These
are amongst some of the subjects to be treated in future studies.

\acknowledgments{We wish to thank A.~Fabbri, R.~Parentani,
S.~Robertson, V.~Rubakov and A.~Vikman for interesting
discussions. The authors acknowledge support from the research
program ``Programme national de cosmologie et galaxies'' of the
CNRS/INSU, France, from the call D\'efi InFIniTi 2017--2018, and
from PRC CNRS/RFBR (2018--2020) No.~1985 ``Modified gravity
and black holes: consistent models and experimental signatures''.}

\appendix
\section{Monopole perturbation}
\label{sec:appmonop}
The aim of this Appendix is to display the explicit expressions of the
various coefficients used in our analysis of Sec.~\ref{sec:effmetrics}.
Those entering Eq.~(\ref{eq:H012Lag}) and later read
\begin{align}
c_1&=-\dfrac{\beta q}{r}\, \varphi' \sqrt{\dfrac{B}{A}},
\\
c_2&=2Bc_1,
\\
c_3&=- \dfrac{1}{2r}\sqrt{\dfrac{B}{A}}\, (-2\zeta A + \beta q^2-3
\beta AB \varphi'^2),
\\
c_4 &= \dfrac{2}{A} c_3,
\\
c_5 &= \dfrac{q^2}{4r^2}\, \dfrac{1}{\sqrt{AB}}
[2\beta(1-B-rB')+\eta r^2],
\\
\begin{split}
c_6 &= -\dfrac{1}{4Ar^2} \sqrt{\dfrac{B}{A}} \left\{\dfrac12 A
\varphi'^2[(2\beta-12 \beta B+\eta r^2)A-12 \beta B r
A']+rA'(\beta q^2-2\zeta A)\right.
\\
&\quad-\left.\vphantom{\dfrac12}2\zeta A^2-\beta q^2 A\right\},
\end{split}
\\
c_7 &= \dfrac{q^2}{4Ar^2}\, \dfrac{1}{\sqrt{AB}} [2 \beta B
rA'+A(2\beta-2\beta B+\eta r^2)],
\\
c_8 &= \dfrac{q}{2Ar^2}\, \varphi' \sqrt{\dfrac{B}{A}} [-6 \beta B
rA'+A(2\beta-6\beta B+\eta r^2)].
\end{align}
The coefficients entering Eq.~(\ref{eq:H0H1Lag}) and later read
\begin{align}
\tilde{c}_1&=-\dfrac{c_1^2}{4c_6},
&\tilde{c}_2&=-\dfrac{c_3^2}{4c_6},
\\
\tilde{c}_3&=\dfrac{c_1 c_3}{2c_6},
&\tilde{c}_4&=-\dfrac{c_4^2}{4c_6},
\\
\tilde{c}_5&=\dfrac{c_4 c_3}{2c_6},
&\tilde{c}_6&=-\dfrac{c_1 c_4}{2c_6},
\\
\tilde{c}_7&=\dfrac{2c_2c_6-c_8c_3}{2c_6},
&\tilde{c}_8&=\dfrac{c_8c_1-c_4c_7}{2c_6},
\\
\tilde{c}_9&=-\dfrac{c_6(c_7c_3'+c_3c_7'-c_7^2+4c_5c_6)-c_7c_3c_6
'}{4c_6^2},
\\
\tilde{c}_{10}&=-\dfrac{c_7 c_8}{2c_6},
&\tilde{c}_{11}&=-\dfrac{c_8^2}{4c_6}.
\end{align}
Finally, the coefficients entering Eq.~(\ref{eq:H01Lag}) read
\begin{align}
a_1 &= \tilde{c}_1,
&a_2 &=\dfrac{\tilde{c}_3}{2\tilde{c}_1},
\\
a_3 &=\dfrac{\tilde{c}_6}{2\tilde{c}_1},
&a_4 &= 0,
\\
a_5 &= \dfrac{2\tilde{c}_1\tilde{c}_7-\tilde{c}_8
\tilde{c}_3}{\tilde{c}_6\tilde{c}_3},
&a_6 &= \dfrac{\tilde{c}_7}{\tilde{c}_3},
\\
a_7 &= \tilde{c}_9-a_1 a_5^2 + (a_1 a_2 a_5)',
&a_8 &= \tilde{c}_{11}-a_1 a_6^2,
\\
a_9 &= \tilde{c}_{10}-2a_1 a_6 a_5.
\end{align}

\end{document}